\documentstyle[12pt,epsfig]{article}
\input{tcilatex}

\begin{document}

\onecolumn

\begin{titlepage}
\begin{center}
{\LARGE \bf Chaos in a 3-body Self-Gravitating Cosmological Spacetime}
\\ \vspace{2cm}
M.J. Koop\footnotemark\footnotetext{email:
mjkoop@sciborg.uwaterloo.ca},
R.B. Mann\footnotemark\footnotetext{email:
rbmann@sciborg.uwaterloo.ca},
\\
\vspace{0.5cm}
Dept. of Physics,
University of Waterloo
Waterloo, ONT N2L 3G1, Canada\\
\vspace{1cm}
S. Bachmann\footnotemark\footnotetext{email:
svenb@phys.ethz.ch}\\
\vspace{0.5cm} Institute for Theoretical Physics, ETH Zurich, 8093 Zurich,
Switzerland \\
\vspace{2cm}
PACS numbers:
13.15.-f, 14.60.Gh, 04.80.+z\\
\vspace{2cm}
\today\\
\end{center}

\begin{abstract}
 We investigate the equal-mass 3-body system in general
relativistic lineal gravity in the presence of a cosmological
constant $\Lambda$. The cosmological vacuum energy introduces
features that do not have a non-relativistic counterpart, inducing
a competing expansion/contraction of spacetime that competes with
the gravitational self-attraction of the bodies. We derive a
canonical expression for the Hamiltonian of the system and discuss
the numerical solution of the resulting equations of motion.  As
for the system with $\Lambda=0$, we find that the structure of the
phase space yields a rich variety of interesting dynamics that
can be divided into three distinct regions: annulus, pretzel, and
chaotic; the first two being regions of quasi-periodicity while
the latter is a region of chaos.  However unlike the $\Lambda=0$
case, we find that a negative cosmological constant considerably
diminishes the amount of chaos in the system,  even beyond that of
the $\Lambda=0$ non-relativistic system.  By contrast, a positive
cosmological constant considerably enhances the amount of chaos,
typically leading to KAM breakdown.
\end{abstract}
\end{titlepage}\onecolumn

\section{INTRODUCTION}

The N-body problem -- that of describing the motion of a system of N
particles interacting through specified forces -- is one of the oldest
problems in physics.\ Even now it continues to be a problem in fields
ranging from nuclear physics to stellar evolution and cosmology.\ When the
problem is that of N particles interacting through their mutual
gravitational attraction it is particularly challenging.\ In the Newtonian
case in three spatial dimensions, an exact solution is only known for the
N=2 case, and for the general relativistic case in three spatial dimensions
there is no exact solution for the equations of motion for even the N=2 case
(although approximation techniques exist \cite{THYoG}). This is largly due
to the dissipation of energy in the form of gravitational radiation, which
results in the system needing to be solved by approximate solutions.

Considerable progress has been made in recent years by studying systems with
reduced spatial dimensions.\ Nonrelativistic one-dimensional
self-gravitating systems (OGS) of N particles have been very important in
the fields of astrophysics and cosmology for over 30 years \cite{Rybicki}.\
Although they are primarily used as prototypes for studying gravity in
higher dimensions, there are physical systems with dynamics closely
approximated by the one dimensional system.\ Very long-lived core-halo
configurations, reminiscent of structures observed in globular clusters, are
known to exist in the OGS phase space \cite{yawn}. These model a dense
massive core in near-equilibrium, surrounded by a halo of high kinetic
energy stars that interact only weakly with the core. Also, the collisions
of flat parallel domain walls moving in a direction perpendicular to their
surfaces and the dynamics of stars in a direction orthogonal to the plane of
a highly flattened galaxy are approximated by the OGS system.

In this paper we continue an ongoing investigation into relativistic
one-dimensional self-gravitating systems (ROGSs). This is carried out in the
context of a (1+1)-dimensional theory of gravity (lineal gravity) that
models (3+1) general relativity by setting the Ricci scalar $R$\ equal to
the trace of the stress energy of prescribed matter fields and sources. \
For the ROGS these sources are point particles minimally coupled to gravity.
Hence, as in (3+1) dimensions, the evolution of space-time curvature is
governed by the matter distribution, which in turn is governed by the
dynamics of space-time \cite{r3}. Referred to as $R=T$\ theory, it is a
particular member of a class of dilaton gravity theories on a line. What
makes this theory of particular interest is that it has a consistent
nonrelativistic ($c\rightarrow \infty $) limit \cite{r3} (a problem for
generic (1+1)-dimensional gravity \cite{jchan}), and in this limit reduces
to the Newtonian OGS as a special case. Furthermore, when the stress energy
is that of a cosmological constant, it reduces to Jackiw-Teitelboim theory %
\cite{JT}.

The three body ROGS has been studied previously and compared to the
corresponding non-relativistic system \cite{3bdshort} in the case of a zero
cosmological constant for both equal and unequal masses \cite{burnell,Justin}%
.\ The degrees of freedom in both the non-relativistic and relativistic
systems can be rewritten in terms of a single particle moving in a two
dimensional potential well of hexagonal symmetry. In the non-relativistic
case the potential grows linearly as a function of radial distance, but in
the relativistic case the growth is non-linear, yielding a potential in the
shape of a hexagonal carafe. For both systems, two broad categories of
periodic and quasi-periodic motion were found, referred to as annulus and
pretzel orbits, as well as a set of chaotic motions appearing in the phase
space between these two types. The phase space between the two systems was
found to be qualitatively the same despite the high degree of nonlinearity
in the relativistic system.

In this paper we study the 3-body ROGS with non-zero cosmological constant $%
\Lambda $. This situation has no non-relativistic analogue. The effect of
the cosmological constant is to induce a competing expansion/contraction of
spacetime that competes with the gravitational self-attraction of the
bodies. This system has been studied in some detail in the case of two
interacting bodies, and it has been shown that relativistic effects are
considerably enhanced by the presence of $\Lambda $ \cite{2bdcoslo}. For
example a sufficiently large and positive $\Lambda $\ can overcome the
attraction of the bodies, which then ''lose casual'' contact over a finite
amount of proper time. The effect of $\Lambda $\ on the chaotic behaviour of
the 3-body system has never been studied. It is the purpose of this paper to
examine this problem, restricting our investigation to the equal-mass case.

In Sec. II we review the formalism of the N-body problem in lineal gravity.
We discuss the canonical decomposition of the action and find the
Hamiltonian of the particles in terms of the dilaton field, which is
determined by a set of constraint equations. We employ two different methods
for solving the constraint equations, which allow us to solve the equations
of motion for systems with both positive and negative values for the
cosmological constant in Sec. III. In Sec. IV we define a coordinate system
that allows us to describe the three particle system as a single particle on
a plane, moving in a potential well with hexagonal symmetry. We describe the
shape of this potential and its dependence on momentum and the cosmological
constant.\ Here we also find upper and lower bounds on allowed values of the
cosmological constant in a system with a given energy. Using these new
coordinates, we describe our methods for numerically solving the system in
Sec. V, and our methods of obtaining orbits, Poincar\'{e} maps, and graphs
of the oscillation patterns of the three particles. In Sec. VI, we
numerically solve the equations of motion in the equal mass case with
different values for the cosmological constant. We find two broad categories
of periodic and quasiperiodic motions that we refer to as annulus and
pretzel patterns, as well as a set of chaotic orbits found in the region of
phase space in between these two types of orbits. We also identify general
changes to the system in the presence of both a positive and a negative
cosmological constant. Finally, in Sec. VII we present various Poincar\'{e}
maps and discuss how the global phase space is distorted in the presence of
a non-zero cosmological constant. We find that a negative cosmological
constant considerably reduces the amount of chaotic behaviour, whereas a
positive cosmological constant considerably enhances it. In Sec. VIII we
discuss the salient features of our solutions and make some conjectures
regarding their general properties. We close the paper with some concluding
remarks and directions for further work.

\bigskip

\section{Canonical Reduction of the N-body Problem in Lineal Gravity}

As in previous work \cite{3bdshort,burnell,Justin} we begin with the action
integral for the gravitational field coupled with N point particles, which is%
\begin{equation}
I=\int d^{2}x\left[ \frac{1}{2\kappa }\sqrt{-g}\left\{ \Psi R+\frac{1}{2}%
g^{\mu \nu }\bigtriangledown _{\mu }\Psi \bigtriangledown _{\nu }\Psi
+\Lambda \right\} -\sum\limits_{a=1}^{N}m_{a}\int d\tau _{a}\left\{ -g_{\mu
\nu }(x)\frac{dz_{a}^{\mu }}{d\tau _{a}}\frac{dz_{a}^{\nu }}{d\tau _{a}}%
\right\} ^{\frac{1}{2}}\delta ^{2}(x-z_{a}(\tau _{a}))\right]
\label{eqn-act1}
\end{equation}%
where $g_{\mu \nu }$ and $g$ are the metric and its determinant, R is the
Ricci scalar, $\tau _{a}$ is the proper time of the $a$-th particle, $\kappa
=8\pi G/c^{4}$ is the gravitational coupling, and with a scalar (dilaton)
field $\Psi $. This action describes a generally covariant self-gravitating
system (without collision terms, so that the bodies pass through each
other), in which the scalar curvature is sourced by the point particles and
the cosmological constant $\Lambda $. Variation of the action with respect
to the metric, particle coordinates, and dilaton field yields the field
equations

\begin{equation}
R-\Lambda =\kappa T_{\mu }^{P\mu }  \label{eqn-rt}
\end{equation}

\begin{equation}
\frac{d}{d\tau _{a}}\left\{ \frac{dz_{a}^{\nu }}{d\tau _{a}}\right\} +\Gamma
_{\alpha \beta }^{\nu }\left( z_{a}\right) \frac{dz_{a}^{\alpha }}{d\tau _{a}%
}\frac{dz_{a}^{\beta }}{d\tau _{a}}=0  \label{eqn-z}
\end{equation}

\begin{equation}
\frac{1}{2}\bigtriangledown _{\mu }\Psi \bigtriangledown _{\nu }\Psi -g_{\mu
\nu }(\frac{1}{4}\bigtriangledown ^{\lambda }\Psi \bigtriangledown _{\lambda
}\Psi -\bigtriangledown ^{2}\Psi )-\bigtriangledown _{\mu }\bigtriangledown
_{\nu }\Psi =\kappa T_{\mu \nu }^{P}+\frac{\Lambda }{2}g_{\mu \nu }
\label{eqn-psi}
\end{equation}%
where the stress-energy due to the point masses is

\begin{equation}
T_{\mu \nu }^{P}=\sum_{a=1}^{N}m_{a}\int d\tau _{a}\frac{1}{\sqrt{-g}}g_{\mu
\sigma }g_{\nu \rho }\frac{dz_{a}^{\sigma }}{d\tau _{a}}\frac{dz_{a}^{\rho }%
}{d\tau _{a}}\delta ^{2}\left( x-z_{a}\left( \tau _{a}\right) \right)
\label{eqn-stressenergy}
\end{equation}%
and is conserved. We observe that (\ref{eqn-rt}) and (\ref{eqn-z}) form a
closed system of N+1 equations for which one can solve for the single metric
degree of freedom and the N degrees of freedom of the point masses. The
evolution of the dilaton field is governed by the evolution of the point
masses via (\ref{eqn-psi}). The left-hand side of (\ref{eqn-psi}) is
divergenceless (consistent with the conservation of $T_{\mu \nu }$),
yielding only one independent equation to determine the single degree of
freedom of the dilaton.

We make use of the decomposition $\sqrt{-g}R=-2\partial _{0}(\sqrt{\gamma }%
K)+2\partial _{1}(\sqrt{\gamma }N^{1}K-\gamma ^{-1}\partial _{1}N_{0})$
where the extrinsic curvature $K=(2N_{0}\gamma )^{-1}(2\partial
_{1}N_{1}-\gamma ^{-1}N_{1}\partial _{1}\gamma -\partial _{0}\gamma )$, and
rewrite the action in the form

\begin{equation}
I=\int dx^{2}\left\{ \sum_{a}p_{a}\dot{z}_{a}\delta \left( x-z_{a}\left(
x^{0}\right) \right) +\pi \dot{\gamma}+\Pi \dot{\Psi}+N_{0}R^{0}+N_{1}R^{1}%
\right\}  \label{eqn-act2}
\end{equation}%
where $\gamma =g_{11}$, $N_{0}=(-g^{00})^{-\frac{1}{2}}$, $N_{1}=g_{10}$, $%
\pi $ and $\Pi $ are conjugate momenta to $\gamma $ and $\Psi $
respectively. The quantities $N_{0}$ and $N_{1}$ are Lagrange multipliers
that enforce the constraints $R^{0}=0=R^{1}$, where%
\begin{equation}
R^{0}=-\kappa \sqrt{\gamma }\gamma \pi ^{2}+2\kappa \sqrt{\gamma }\pi \Pi +%
\frac{\left( \Psi ^{\prime }\right) ^{2}}{4\kappa \sqrt{\gamma }}-\left( 
\frac{\Psi ^{\prime }}{\kappa \sqrt{\gamma }}\right) ^{\prime }+\frac{%
\Lambda }{2\kappa }\sqrt{\gamma }-\sum_{a}\sqrt{\frac{p_{a}^{2}}{\gamma }%
+m_{a}^{2}}\delta \left( x-z_{a}\left( x^{0}\right) \right)  \label{eqn-R0}
\end{equation}%
\begin{equation}
R^{1}=\frac{\gamma ^{\prime }}{\gamma }\pi -\frac{1}{\gamma }\Pi \Psi
^{\prime }+2\pi ^{\prime }+\sum_{a}\frac{p_{a}}{\gamma }\delta \left(
x-z_{a}\left( x^{0}\right) \right)  \label{eqn-R1}
\end{equation}%
with the symbols ( $^{\cdot }$\ ) and ( $^{\prime }$ ) denoting $\partial
_{0}$ and $\partial _{1}$, respectively. This action leads to the following
system of field equations:

{\it 
\begin{eqnarray}
&&\dot{\pi}+N_{0}\left\{ \frac{3\kappa }{2}\sqrt{\gamma }\pi ^{2}-\frac{%
\kappa }{\sqrt{\gamma }}\pi \Pi +\frac{1}{8\kappa \sqrt{\gamma }\gamma }%
\left( \Psi ^{\prime }\right) ^{2}-\frac{\Lambda }{4\kappa \sqrt{\gamma }}%
-\sum_{a}\frac{p_{a}^{2}}{2\gamma ^{2}\sqrt{\frac{p_{a}^{2}}{\gamma }%
+m_{a}^{2}}}\delta \left( x-z_{a}\left( t\right) \right) \right\}  \nonumber
\\
&&+N_{1}\left\{ -\frac{1}{\gamma ^{2}}\Pi \Psi ^{\prime }+\frac{\pi ^{\prime
}}{\gamma }+\sum_{a}\frac{p_{a}}{\gamma ^{2}}\delta \left( x-z_{a}\left(
t\right) \right) \right\} +N_{0}^{\prime }\frac{1}{2\kappa \sqrt{\gamma }%
\gamma }\Psi ^{\prime }+N_{1}^{\prime }\frac{\pi }{\gamma }=0  \label{feq1}
\end{eqnarray}%
}

{\it 
\begin{equation}
\dot{\gamma}-N_{0}\left( 2\kappa \sqrt{\gamma }\gamma \pi -2\kappa \sqrt{%
\gamma }\Pi \right) +N_{1}\frac{\gamma ^{\prime }}{\gamma }-2N_{1}^{\prime
}=0  \label{feq2}
\end{equation}%
}

{\it 
\begin{equation}
R^{0}=0  \label{feq3}
\end{equation}%
}

{\it 
\begin{equation}
R^{1}=0  \label{feq4}
\end{equation}%
}

{\it 
\begin{equation}
\dot{\Pi}+\partial _{1}\left( -\frac{1}{\gamma }N_{1}\Pi +\frac{1}{2\kappa 
\sqrt{\gamma }}N_{0}\Psi ^{\prime }+\frac{1}{\kappa \sqrt{\gamma }}%
N_{0}^{\prime }\right) =0  \label{feq5}
\end{equation}%
}

{\it 
\begin{equation}
\dot{\Psi}+N_{0}\left( 2\kappa \sqrt{\gamma }\pi \right) -N_{1}\left( \frac{1%
}{\gamma }\Psi ^{\prime }\right) =0  \label{feq6}
\end{equation}%
}

{\it 
\begin{equation}
\dot{p}_{a}+\frac{\partial N_{0}}{\partial z_{a}}\sqrt{\frac{p_{a}^{2}}{%
\gamma }+m_{a}^{2}}-\frac{N_{0}}{2\sqrt{\frac{p_{a}^{2}}{\gamma }+m_{a}^{2}}}%
\frac{p_{a}^{2}}{\gamma ^{2}}\frac{\partial \gamma }{\partial z_{a}}-\frac{%
\partial N_{1}}{\partial z_{a}}\frac{p_{a}}{\gamma }+N_{1}\frac{p_{a}}{%
\gamma ^{2}}\frac{\partial \gamma }{\partial z_{a}}=0  \label{feq7}
\end{equation}%
}

{\it 
\begin{equation}
\dot{z}_{a}-N_{0}\frac{\frac{p_{a}}{\gamma }}{\sqrt{\frac{p_{a}^{2}}{\gamma }%
+m_{a}^{2}}}+\frac{N_{1}}{\gamma }=0  \label{feq8}
\end{equation}%
}

We can solve the constraint equations (\ref{feq3}) and (\ref{feq4}) in terms
of the quantities ($\Psi ^{\prime }/\sqrt{\gamma })^{\prime }$\ and $\pi
^{\prime }$, since they are the only linear terms present. The generator
obtained from the end point variation can then be transformed to fix the
coordinate conditions. We can consistently choose the coordinate conditions $%
\gamma =1$\ and $\Pi =0$\ \cite{OR}. Eliminating the constraints, the action
then reduces to 
\begin{equation}
I=\int d^{2}x\left\{ \sum_{a}p_{a}\dot{z}_{a}\delta \left( x-z_{a}\right) -%
{\cal H}\right\}  \label{eqn-act3}
\end{equation}%
where the reduced Hamiltonian is%
\begin{equation}
H=\int dx{\cal H}=-\frac{1}{\kappa }\int dx\Delta \Psi  \label{Ham1}
\end{equation}%
where $\Delta \equiv \partial ^{2}/\partial x^{2}$, and $\Psi =\Psi \left(
x,z_{a},p_{a}\right) $ and is understood to be determined from the
constraint equations which are now

\begin{equation}
\Delta \Psi -\frac{\left( \Psi ^{\prime }\right) ^{2}}{4}+\kappa ^{2}\pi
^{2}-\frac{\Lambda }{2}+\kappa \sum_{a}\sqrt{p_{a}^{2}+m_{a}^{2}}\delta
\left( x-z_{a}\right) =0  \label{cst1}
\end{equation}%
\begin{equation}
2\Delta \chi +\sum_{a}p_{a}\delta \left( x-z_{a}\right) =0  \label{cst2}
\end{equation}%
where $\pi =\chi ^{\prime }$. The consistency of this canonical reduction
can be demonstrated by showing that the canonical equations of motion
derived from the reduced Hamiltonian (\ref{Ham1}) are identical with (\ref%
{feq7}) and (\ref{feq8}) \cite{OR}.

\bigskip

\section{Solving the Constraint Equations}

We can solve these two equations exactly for $N=3$ particles by solving the
equations in the empty spaces between particles and then matching the fields
at each of the particle positions, imposing the condition that the
Hamiltonian remains finite as $x\rightarrow \infty $. After these
constraints are applied we are left with an equation containing only one
unknown.

We divide the space into four regions determined by the arbitrary particle
positions chosen so that $z_{1}<z_{2}<z_{3}$. We label the four regions: $%
x<z_{1}$ ((1) region), $z_{1}<x<z_{2}$, ((2) region), $z_{2}<x<z_{3}$ ((3)
region), and $z_{3}<x$ ((4) region). In each of these regions the solution
to (\ref{cst2}) is simply

\begin{equation}
\chi =-\frac{1}{4}\sum_{a=1}^{3}p_{a}\left| x-z_{a}\right| -\epsilon
Xx+\epsilon C_{\chi }  \label{eqn-Ki}
\end{equation}%
with $X$ and $C_{\chi }$ being two integration constants. The factor $%
\epsilon $\ ($\epsilon ^{2}=1$) changes sign under time reversal; it has
been introduced to explicitly manifest the property that $\chi $\ also
changes sign under time reversal. We find an easier expression for (\ref%
{cst1}) by using the substitution%
\begin{equation}
\Psi =-4\log \left| \phi \right|  \label{Psi-sub}
\end{equation}%
which gives us%
\begin{equation}
\Delta \phi -\frac{1}{4}\phi \left\{ -\frac{\Lambda }{2}+\kappa ^{2}\left(
\chi ^{\prime }\right) ^{2}\right\} =\frac{\kappa }{4}\sum_{a=1}^{3}\sqrt{%
p_{a}^{2}+m_{a}^{2}}\phi \left( z_{a}\right) \delta (x-z_{a})
\label{cst1-phi}
\end{equation}

In the four regions defined above equation (\ref{cst1-phi}) has the solution%
\begin{equation}
\phi =\left\{ 
\begin{array}{cc}
A_{1}e^{\frac{1}{2}K_{1}x}+B_{1}e^{-\frac{1}{2}K_{1}x} & \text{in the (1)
region} \\ 
A_{2}e^{\frac{1}{2}K_{2}x}+B_{2}e^{-\frac{1}{2}K_{2}x} & \text{in the (2)
region} \\ 
A_{3}e^{\frac{1}{2}K_{3}x}+B_{3}e^{-\frac{1}{2}K_{3}x} & \text{in the (3)
region} \\ 
A_{4}e^{\frac{1}{2}K_{4}x}+B_{4}e^{-\frac{1}{2}K_{4}x} & \text{in the (4)
region}%
\end{array}%
\right\}  \label{eqn-phi}
\end{equation}%
where the $K_{i}$ are defined as%
\begin{equation}
\left\{ 
\begin{array}{c}
K_{1}=\sqrt{\kappa ^{2}\left[ X-\frac{\epsilon }{4}(p_{1}+p_{2}+p_{3})\right]
^{2}-\frac{\Lambda }{2}} \\ 
K_{2}=\sqrt{\kappa ^{2}\left[ X-\frac{\epsilon }{4}(-p_{1}+p_{2}+p_{3})%
\right] ^{2}-\frac{\Lambda }{2}} \\ 
K_{3}=\sqrt{\kappa ^{2}\left[ X+\frac{\epsilon }{4}(p_{1}+p_{2}-p_{3})\right]
^{2}-\frac{\Lambda }{2}} \\ 
K_{4}=\sqrt{\kappa ^{2}\left[ X+\frac{\epsilon }{4}\left(
p_{1}+p_{2}+p_{3}\right) \right] ^{2}-\frac{\Lambda }{2}}%
\end{array}%
\right\}  \label{Kdef}
\end{equation}%
The matching conditions that must be satisfied require that $\phi $ is
continuous and its derivatives are consistent with integration of eq. (\ref%
{cst1-phi}) across the location of each particle: 
\begin{eqnarray}
\phi _{2}\left( z_{1}^{+}\right) &=&\phi _{1}\left( z_{1}^{-}\right)
\label{eqn-BC} \\
\phi _{3}\left( z_{2}^{+}\right) &=&\phi _{2}\left( z_{2}^{-}\right) 
\nonumber \\
\phi _{4}\left( z_{3}^{+}\right) &=&\phi _{3}\left( z_{3}^{-}\right) 
\nonumber \\
\phi _{2}^{\prime }\left( z_{1}^{+}\right) -\phi _{1}^{\prime }\left(
z_{1}^{-}\right) &=&\frac{\kappa }{4}\sqrt{p_{1}^{2}+m_{1}^{2}}\phi \left(
z_{1}\right)  \nonumber \\
\phi _{3}^{\prime }\left( z_{2}^{+}\right) -\phi _{2}^{\prime }\left(
z_{2}^{-}\right) &=&\frac{\kappa }{4}\sqrt{p_{2}^{2}+m_{2}^{2}}\phi \left(
z_{2}\right)  \nonumber \\
\phi _{4}^{\prime }\left( z_{3}^{+}\right) -\phi _{3}^{\prime }\left(
z_{3}^{-}\right) &=&\frac{\kappa }{4}\sqrt{p_{3}^{2}+m_{3}^{2}}\phi \left(
z_{3}\right)  \nonumber
\end{eqnarray}%
where $\phi _{a}\left( z_{\beta }^{+}\right) =\lim_{x\rightarrow z_{\beta
},x>z_{\beta }}\phi _{\alpha }\left( x\right) $.

This leaves us with six equations for ten unknowns ($A$'s, $B$'s, $X$, and $%
C_{\chi }$). There is also the condition that the Hamiltonian remains
finite, which can be done by requiring%
\begin{equation}
\Psi ^{2}-4\kappa ^{2}\chi ^{2}+2\Lambda x^{2}=C_{\pm }x  \label{eqn-BC2}
\end{equation}%
where $C_{\pm }$ are constants to be determined \cite{OR,2bdcossh}. This
gives%
\[
A_{1}=B_{4}=0 
\]%
and four other equations that contain $C_{\pm }$.\ This leaves us with ten
equations for ten unknowns ($A_{2}$, $A_{3}$, $A_{4}$, $B_{1}$, $B_{2}$, $%
B_{3}$, $C_{\chi }$, $C_{-}$, $C_{+}$, $X$).\ At this point, one can do the
integration explicitly.\ The Hamiltonian becomes%
\[
H=-\frac{1}{\kappa }\int dx\Delta \Psi =-\frac{1}{\kappa }\left[ -\frac{%
4\phi ^{\prime }}{\phi }\right] _{-\infty }^{+\infty }=\frac{2\left(
K_{1}+K_{4}\right) }{\kappa } 
\]%
Without loss of generality, we choose the centre of inertia frame $%
p_{1}+p_{2}+p_{3}=0$, which simplifies the Hamiltonian to%
\begin{equation}
H=\frac{4}{\kappa }\sqrt{\kappa ^{2}X^{2}-\frac{\Lambda }{2}}  \label{Ham2}
\end{equation}%
The Hamiltonian can only depend upon the relative separation of the
particles and their conjugate momenta. Furthermore, in the equal mass case,
the system is symmetric under particle interchange. We introduce the
following notation%
\begin{equation}
z_{ij}\equiv (z_{i}-z_{j})  \label{delta-z}
\end{equation}%
\begin{equation}
s_{ij}\equiv \text{sgn}\left( z_{ij}\right)  \label{eqn-sij}
\end{equation}%
\begin{equation}
R_{i\pm }\equiv \sqrt{\kappa ^{2}\left[ X-\frac{\epsilon }{4}\left(
\sum_{a=1}^{3}p_{a}s_{ai}\pm p_{i}\right) \right] ^{2}-\frac{\Lambda }{2}}
\label{eqn-Rpm}
\end{equation}%
\begin{equation}
M_{ij}\equiv \kappa \sqrt{p_{i}^{2}+m_{i}^{2}}+2s_{ij}\left(
R_{i+}-R_{i-}\right)  \label{eqn-Mij}
\end{equation}%
\begin{equation}
L_{i}\equiv -\kappa \sqrt{p_{i}^{2}+m_{i}^{2}}+2\left( R_{i+}+R_{i-}\right)
\label{eqn-Li}
\end{equation}%
\begin{equation}
L_{i}^{\ast }\equiv \left( -\sum_{j<k\neq i}s_{ij}s_{ik}\right) \kappa \sqrt{%
p_{i}^{2}+m_{i}^{2}}+2\left( R_{i+}+R_{i-}\right)  \label{eqn-Lstar}
\end{equation}%
where these quantities are defined in such a way that they automatically
take care of the crossings of particles via $s_{ij}$. Using them, the
derivation of the full determining equation for the Hamiltonian is similar
to that for the $\Lambda =0$ case \cite{burnell}; the result is

\begin{eqnarray}
L_{1}L_{2}L_{3} &=&M_{12}M_{21}L_{3}^{\ast }e^{\frac{1}{4}s_{21}\left[
\left( M_{12}+L_{1}\right) z_{31}-\left( M_{21}+L_{2}\right) z_{32}\right] }
\label{det-eqn1} \\
&&+M_{23}M_{32}L_{1}^{\ast }e^{\frac{1}{4}s_{32}\left[ \left(
M_{23}+L_{2}\right) z_{12}-\left( M_{32}+L_{3}\right) z_{13}\right] } 
\nonumber \\
&&+M_{31}M_{13}L_{2}^{\ast }e^{\frac{1}{4}s_{13}\left[ \left(
M_{31}+L_{3}\right) z_{23}-\left( M_{13}+L_{1}\right) z_{21}\right] } 
\nonumber
\end{eqnarray}%
or more compactly%
\begin{equation}
L_{1}L_{2}L_{3}=\sum_{ijk}|\epsilon _{ijk}|M_{ij}M_{ji}L_{k}^{\ast }e^{\frac{%
1}{4}s_{ij}\left[ \left( M_{ij}+L_{i}\right) z_{ki}-\left(
M_{ji}+L_{j}\right) z_{kj}\right] }  \label{det-eqn2}
\end{equation}%
where $\epsilon _{ijk}$ is the Levi-Civita tensor.

While technically correct, the determining equation (\ref{det-eqn1}) is not
useful computationally insofar as some of its terms can become imaginary
with a sufficiently large cosmological constant; this is a consequence of
eq. (\ref{Kdef}). However we can rewrite the determining equation to
circumvent this problem. Consider the case where $z_{1}<z_{2}<z_{3}$\ .
Redefining the following terms%
\begin{eqnarray}
\hat{M}_{i} &=&\kappa \sqrt{p_{i}^{2}+m_{i}^{2}}  \label{new_def} \\
\hat{K}_{j} &=&-2\sqrt{\kappa ^{2}\left[ X+\frac{\epsilon }{4}\left(
\sum_{i=1}^{3}\sigma _{ji}p_{i}\right) \right] ^{2}-\frac{\Lambda }{2}} 
\nonumber
\end{eqnarray}%
where $\sigma _{ji}=-1$ when $j\leq i$ and $\sigma _{ji}=1$ when $i<j$ we
can get a new form for the determining equation 
\begin{eqnarray}
&&\left[ \left( \left( \hat{M}_{1}+\hat{K}_{1}\right) \left( \hat{M}_{3}+%
\hat{K}_{4}\right) \hat{M}_{2}+\left( \hat{M}_{1}+\hat{K}_{1}\right) \hat{K}%
_{3}^{2}+\left( \hat{M}_{3}+\hat{K}_{4}\right) \hat{K}_{2}^{2}\right) \tanh
\left( \frac{1}{4}\hat{K}_{3}z_{32}\right) \tanh \left( \frac{1}{4}\hat{K}%
_{2}z_{21}\right) \right.  \nonumber \\
&&+\left( \left( \hat{M}_{1}+\hat{M}_{2}+\hat{K}_{1}\right) \left( \hat{M}%
_{3}+\hat{K}_{4}\right) +\hat{K}_{3}^{2}\right) \hat{K}_{2}\tanh \left( 
\frac{1}{4}\hat{K}_{3}z_{32}\right)  \nonumber \\
&&+\left( \left( \hat{M}_{1}+\hat{K}_{1}\right) \left( \hat{M}_{2}+\hat{M}%
_{3}+\hat{K}_{4}\right) +\hat{K}_{2}^{2}\right) \hat{K}_{3}\tanh \left( 
\frac{1}{4}\hat{K}_{2}z_{21}\right)  \nonumber \\
&&\left. +\left( \hat{M}_{1}+\hat{M}_{2}+\hat{M}_{3}+\hat{K}_{1}+\hat{K}%
_{4}\right) \hat{K}_{2}\hat{K}_{3}\right]  \nonumber \\
&=&0  \label{det-eqn3}
\end{eqnarray}%
When the center of momentum is set to zero only the terms $\hat{K}_{2}$ or $%
\hat{K}_{3}$ can be imaginary. Consequently (\ref{det-eqn3})\ is either
purely real, or purely imaginary (in which case the $i$ can be factored
out). \ Permutation of the particles results in the exact same determining
equation with the indicies appropriately switched.

The Hamiltonian $H$\ is only implicitly determined from (\ref{det-eqn3}).
Fortunately we do not need its explicit form since we can implicitly take
derivatives on both sides of the determining equation (\ref{det-eqn3}) with
respect to the phase space variables and from there extract the canonical
equations of motion. From (\ref{Ham2}) with $p_{Z}=\sum_{a=1}^{3}p_{a}=0$,
we obtain%
\begin{equation}
\frac{\partial H}{\partial x_{k}}=\frac{4}{\kappa }\left( \frac{\partial }{%
\partial x_{k}}\sqrt{\kappa ^{2}X^{2}-\frac{\Lambda }{2}}\right) =\frac{%
4\kappa X}{\sqrt{\kappa ^{2}X^{2}-\frac{\Lambda }{2}}}\frac{\partial X}{%
\partial x_{k}}  \label{eqn-dHtodX}
\end{equation}%
where $x_{k}$\ is any of the canonical variables. \ We then have the simple
but tedious task of taking $\frac{\partial }{\partial x_{k}}$\ of both sides
of equation (\ref{det-eqn3}), collecting the derivatives of $X$, and then
converting them to derivatives of $H$ via (\ref{eqn-dHtodX}). This yields
the canonical equations of motion%
\begin{equation}
\dot{z}_{a}=\frac{\partial H}{\partial p_{a}}  \label{CanonEqn1}
\end{equation}%
\begin{equation}
\dot{p}_{a}=-\frac{\partial H}{\partial z_{a}}  \label{CanonEqn2}
\end{equation}%
where the dot denotes a derivative with respect to the coordinate time.
These equations of motion are straightforward to calculate but highly
tedious and will not be included here. When calculating the equations of
motion for the particles when they are not in the arrangement $%
z_{1}<z_{2}<z_{3}$\ we temporarily change the labels of the particles so
that they do satisfy this condition. This allows us to use the derivatives
of (\ref{det-eqn3}) to calculate the equations of motion for each of the
particles, after which the original particle labels are returned. This
method works regardless of whether or not the particles are identical.

\section{General Properties of the Equations of Motion}

We can now numerically calculate the equations of motion for each of the
three particles. This would give us six equations of motion when our
physical system only has four effective degrees of\ freedom: two relative
separations and their two conjugate momenta. \ Introducing the new variables

\begin{eqnarray}
z_{2}-z_{1} &=&\sqrt{2}\rho  \label{eqn-rholambda} \\
-\left( z_{1}+z_{2}\right) +2z_{3} &=&\sqrt{6}\lambda  \nonumber \\
z_{1}+z_{2}+z_{3} &=&Z  \nonumber
\end{eqnarray}%
and their conjugate momenta%
\begin{eqnarray}
p_{\rho } &=&\frac{1}{\sqrt{2}}(p_{2}-p_{1})  \label{eqn-prhoplambda} \\
p_{\lambda } &=&\frac{1}{\sqrt{6}}\left( -\left( p_{1}+p_{2}\right)
+2p_{3}\right)  \nonumber \\
p_{Z} &=&\frac{1}{3}\left( p_{1}+p_{2}+p_{3}\right)  \nonumber
\end{eqnarray}%
which have been chosen to satisfy the canonical commutation relations $%
\left\{ q_{i},p_{j}\right\} =\delta _{ij}$. This choice of variables gives
the Hamiltonian an explicit sixfold symmetry. We can set the conjugate
variables $Z$ and $p_{Z}$ to arbitrary values since $Z$ is irrelevant in our
equations and we can fix the center of inertia by setting $p_{Z}=0$ without
loss of generality. Inverting these relations, we get%
\begin{eqnarray}
z_{2}-z_{1} &=&\sqrt{2}\rho  \label{eqn-zdiff} \\
z_{3}-z_{2} &=&\frac{1}{\sqrt{2}}(\sqrt{3}\lambda -\rho )  \nonumber \\
z_{3}-z_{1} &=&\frac{1}{\sqrt{2}}(\sqrt{3}\lambda +\rho )  \nonumber
\end{eqnarray}%
and%
\begin{eqnarray}
p_{1} &=&-\frac{1}{\sqrt{2}}p_{\rho }-\frac{1}{\sqrt{6}}p_{\lambda }
\label{eqn-p} \\
p_{2} &=&\frac{1}{\sqrt{2}}p_{\rho }-\frac{1}{\sqrt{6}}p_{\lambda } 
\nonumber \\
p_{3} &=&\sqrt{\frac{2}{3}}p_{\lambda }  \nonumber
\end{eqnarray}

This choice of variables allows us to write the determining equation in
terms of only $\left\{ \rho ,\lambda ,p_{\rho },p_{\lambda }\right\} $ for a
given cosmological constant.\ In this way, the relativistic Hamiltonian can
be regarded as a function $H=H\left( \rho ,\lambda ,p_{\rho },p_{\lambda
}\right) $

We see from this perspective that the linear three-particle system is
equivalent to a single particle moving in a two dimensional ``potential
well''. This allows the variables $\rho ,\lambda ,p_{\rho },p_{\lambda }$ to
be interpreted as the coordinates and the conjugate momenta of this single
particle, which we call the hex-particle due to the hexagonal symmetry of
the potential. In the Newtonian case, the potential takes on the shape of a
hexagonal-shaped cone with planar sides and is independent of the momenta %
\cite{Bukta}.\thinspace The relativistic potential well is obtained by
regarding the potential to be the difference between the Hamiltonian and the
relativistic kinetic energy 
\begin{equation}
V\left( \rho ,\lambda \right) |_{p_{\rho }=a,p_{\lambda }=b}=H\left( \rho
,\lambda ,p_{\rho }=a,p_{\lambda }=b\right) -\sqrt{\left( mc^{2}\right)
^{2}+\left( c\left| p\right| \right) ^{2}}  \label{eqn-potential}
\end{equation}%
\ and is dependent on the momentum $\left| p\right| =\sqrt{a^{2}+b^{2}}$ of
the hex-particle as well as its position in the $\left( \rho ,\lambda
\right) $\ plane\footnote{{\it Note that earlier definitions of the
potential \cite{burnell} defined it to be the value of the Hamiltonian at
zero momenta.}}.

\bigskip

\bigskip We get our first look at the relativistic potential by considering
the case when $p_{\rho }=0=p_{\lambda }$:%
\[
V\left( \rho ,\lambda \right) |_{p_{\rho }=0,p_{\lambda }=0}=H\left( \rho
,\lambda ,p_{\rho }=0,p_{\lambda }=0\right) -mc^{2} 
\]%
At very low energies this relativistic potential is indistinguishable from
the potential for the Newtonian case. However, as shown in table (1), even
at energies only moderately larger than the rest mass, the sides of the
hexagon become convex in the relativistic potential.\ Though it retains its
hexagonal symmetry, the growth of the relativistic potential is very rapid
as $\rho $ and $\lambda $ increase. The size of the cross section of the
potential reaches a maximum at an energy $V_{Rc}$ just over twice the rest
energy of the particle, after which the diameter of the potential decreases
like $\ln \left( V_{R}\right) /V_{R}$ with increasing $V_{R}$ \cite{burnell}%
. \ 

The part of the potential on the branch with $V_{R}>V_{Rc}$ is in an
intrinsically nonperturbative relativistic regime.\ The motion for values of 
$V_{R}$ larger than this cannot be understood as a perturbation from some
classical limit of the motion. The nonrelativistic hexagonal cone becomes a
hexagonal carafe in the relativistic case, with a neck that narrows as $%
V_{R} $ increases.

When $p_{\rho }=0=p_{\lambda }$\ it is straightforward to show that the
potential is independent of the cosmological constant, a feature noted
previously in the 2-body system \cite{2bdcoslo,2bdcossh}. However, for
nonzero $p$\ this is not the case, and we map out the potential\ to show how
it changes as $\Lambda $\ varies. It is convenient to instead map the
potential with set values of radial momentum $p_{r}$\ and angular momentum $%
p_{\theta }$\ of the hex-particle where

{\it 
\[
p_{r}=p_{\rho }\cos \theta +p_{\lambda }\sin \theta 
\]%
}

{\it 
\[
p_{\theta }=-p_{\rho }\sin \theta +p_{\lambda }\cos \theta 
\]%
}

\[
\theta =\arctan \left( \frac{\lambda }{\rho }\right) 
\]%
since the change to polar coordinates manifestly retains the hexagonal
symmetry of the potential, which greatly simplifies the subsequent analysis.

%TCIMACRO{\TeXButton{B}{\begin{table}[tbp] \centering}}%
%BeginExpansion
\begin{table}[tbp] \centering%
%EndExpansion
\begin{tabular}{|ll|}
\hline
\FRAME{itbpFU}{2.0358in}{2.0358in}{0in}{\Qcb{$\Lambda =0$ \ \ \ $p_{r}=0.3$}%
}{}{zl_pp_side.eps}{\special{language "Scientific Word";type
"GRAPHIC";maintain-aspect-ratio TRUE;display "USEDEF";valid_file "F";width
2.0358in;height 2.0358in;depth 0in;original-width 7.9442in;original-height
7.9442in;cropleft "0";croptop "1";cropright "1";cropbottom "0";filename
'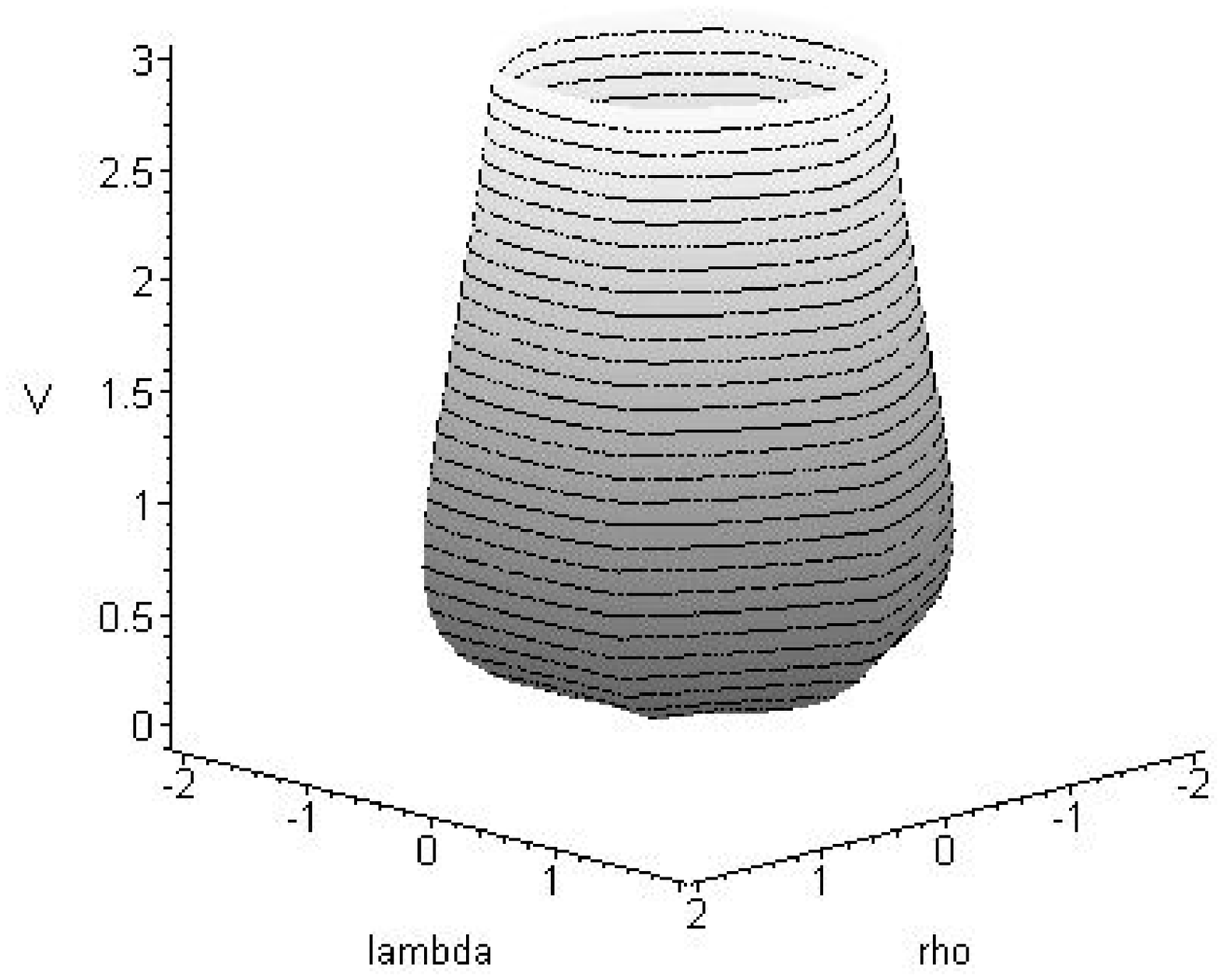';file-properties "XNPEU";}} & \FRAME{itbpF}{2.0358in}{%
2.0358in}{0in}{}{}{zl_pp_top.eps}{\special{language "Scientific Word";type
"GRAPHIC";maintain-aspect-ratio TRUE;display "USEDEF";valid_file "F";width
2.0358in;height 2.0358in;depth 0in;original-width 7.9442in;original-height
7.9442in;cropleft "0";croptop "1";cropright "1";cropbottom "0";filename
'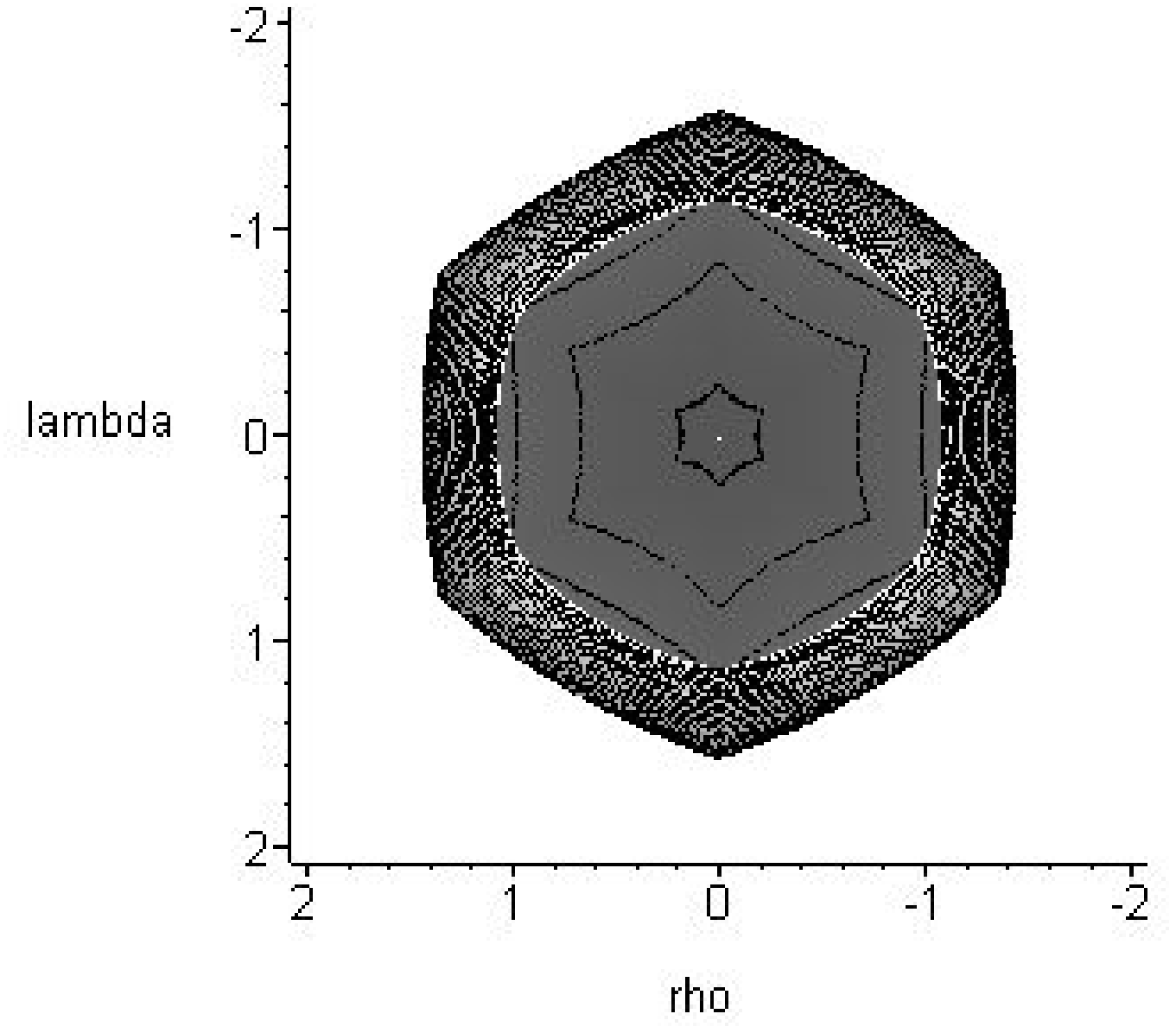';file-properties "XNPEU";}} \\ \hline
\FRAME{itbpFU}{2.0358in}{2.0358in}{0in}{\Qcb{$\Lambda =0$ \ \ \ $p_{r}=0$}}{%
}{zl_zp_side.eps}{\special{language "Scientific Word";type
"GRAPHIC";maintain-aspect-ratio TRUE;display "USEDEF";valid_file "F";width
2.0358in;height 2.0358in;depth 0in;original-width 7.9442in;original-height
7.9442in;cropleft "0";croptop "1";cropright "1";cropbottom "0";filename
'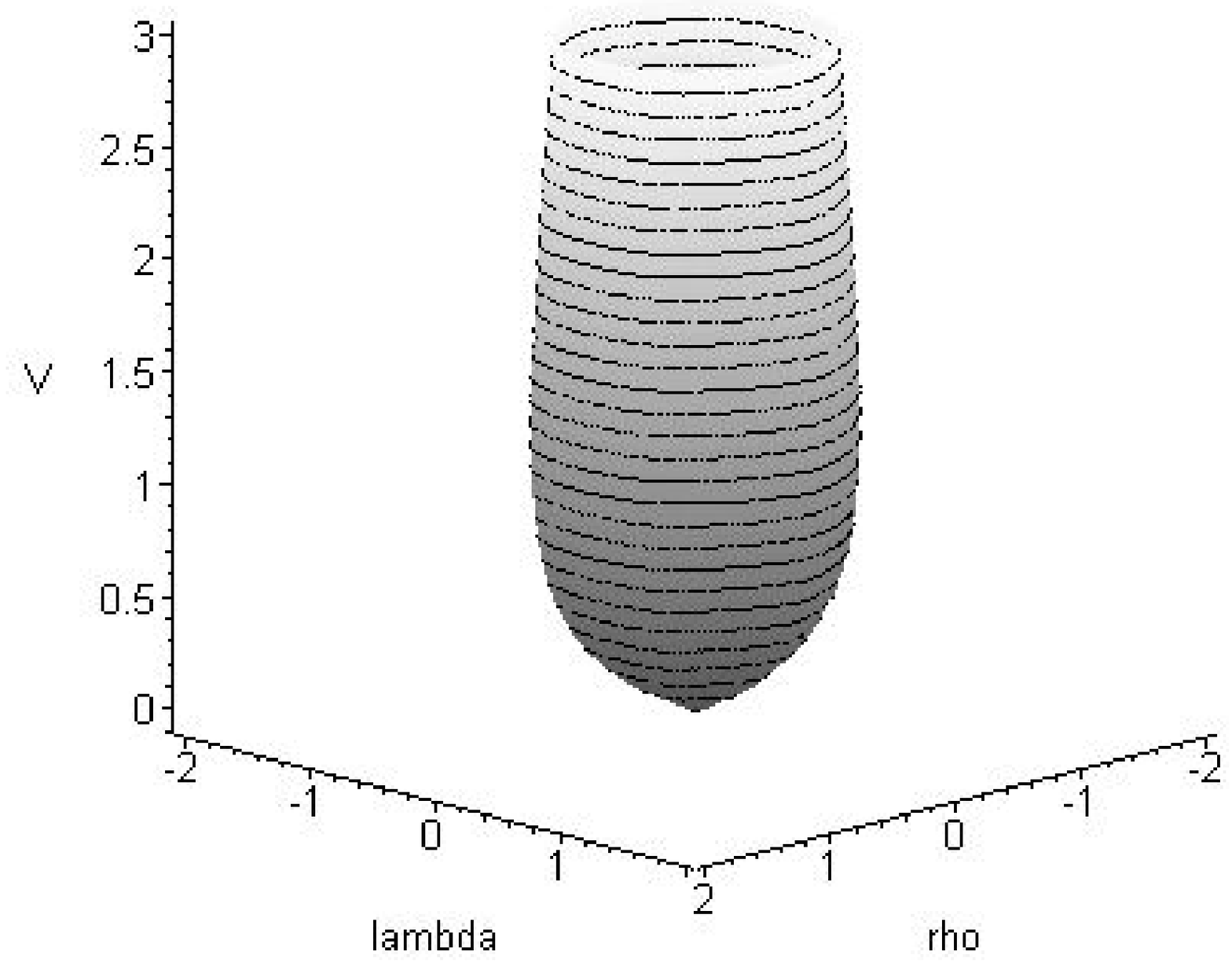';file-properties "XNPEU";}} & \FRAME{itbpF}{2.0358in}{%
2.0358in}{0in}{}{}{zl_zp_top.eps}{\special{language "Scientific Word";type
"GRAPHIC";maintain-aspect-ratio TRUE;display "USEDEF";valid_file "F";width
2.0358in;height 2.0358in;depth 0in;original-width 7.9442in;original-height
7.9442in;cropleft "0";croptop "1";cropright "1";cropbottom "0";filename
'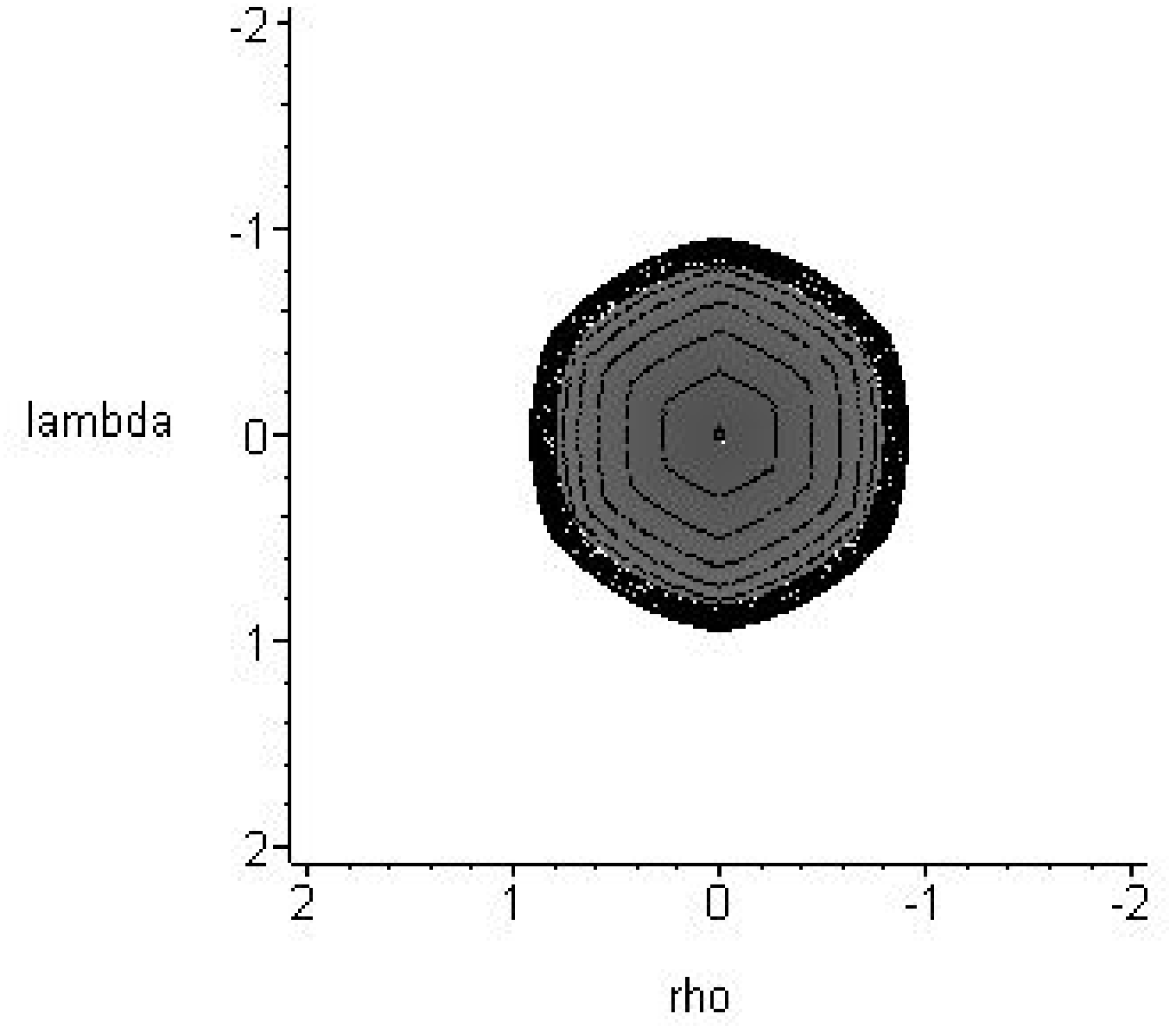';file-properties "XNPEU";}} \\ \hline
\FRAME{itbpFU}{2.0358in}{2.0358in}{0in}{\Qcb{$\Lambda =0$ \ \ \ $p_{r}=-0.3$}%
}{}{zl_np_side.eps}{\special{language "Scientific Word";type
"GRAPHIC";maintain-aspect-ratio TRUE;display "USEDEF";valid_file "F";width
2.0358in;height 2.0358in;depth 0in;original-width 7.9442in;original-height
7.9442in;cropleft "0";croptop "1";cropright "1";cropbottom "0";filename
'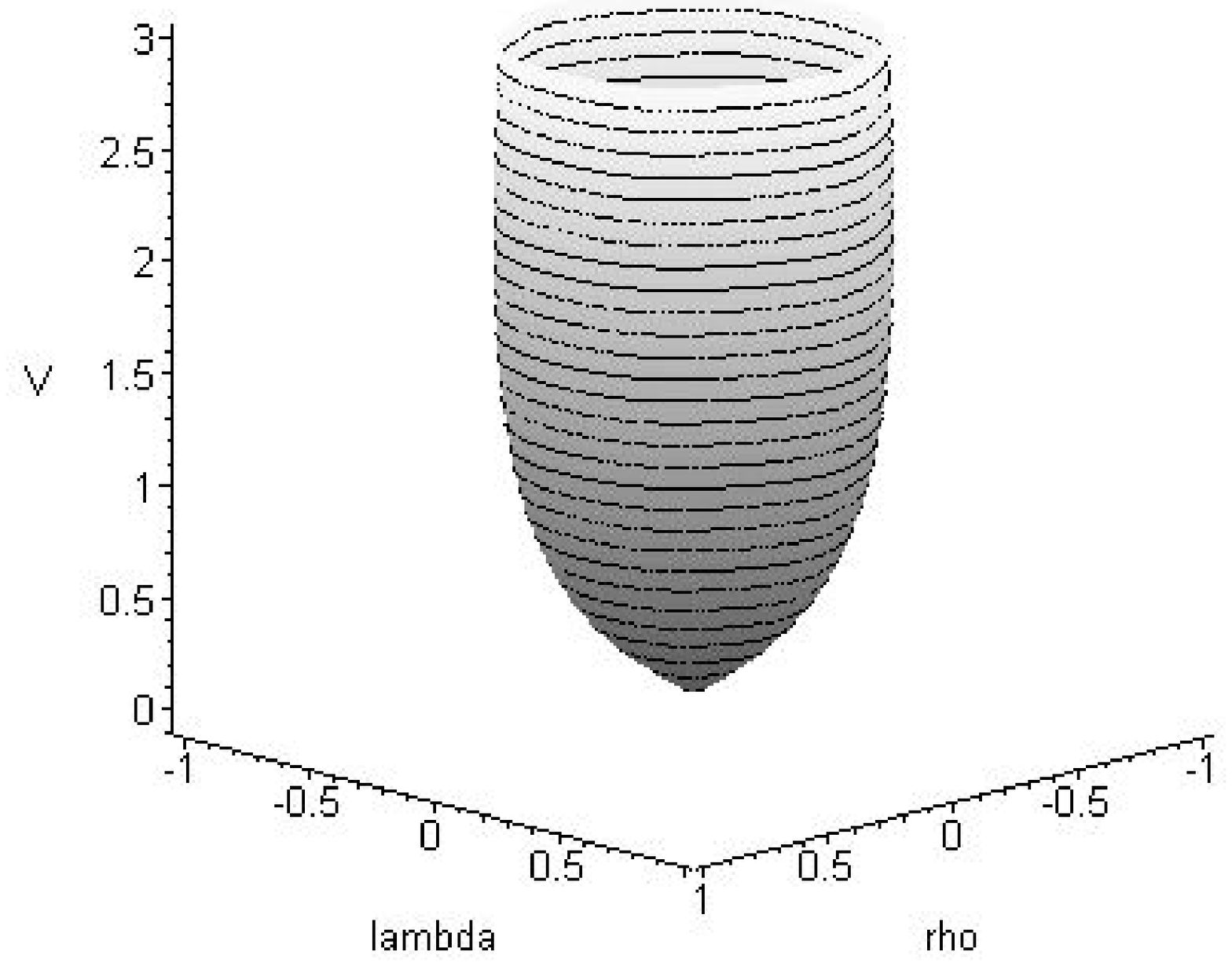';file-properties "XNPEU";}} & \FRAME{itbpF}{2.0358in}{%
2.0358in}{0in}{}{}{zl_np_top.eps}{\special{language "Scientific Word";type
"GRAPHIC";maintain-aspect-ratio TRUE;display "USEDEF";valid_file "F";width
2.0358in;height 2.0358in;depth 0in;original-width 7.9442in;original-height
7.9442in;cropleft "0";croptop "1";cropright "1";cropbottom "0";filename
'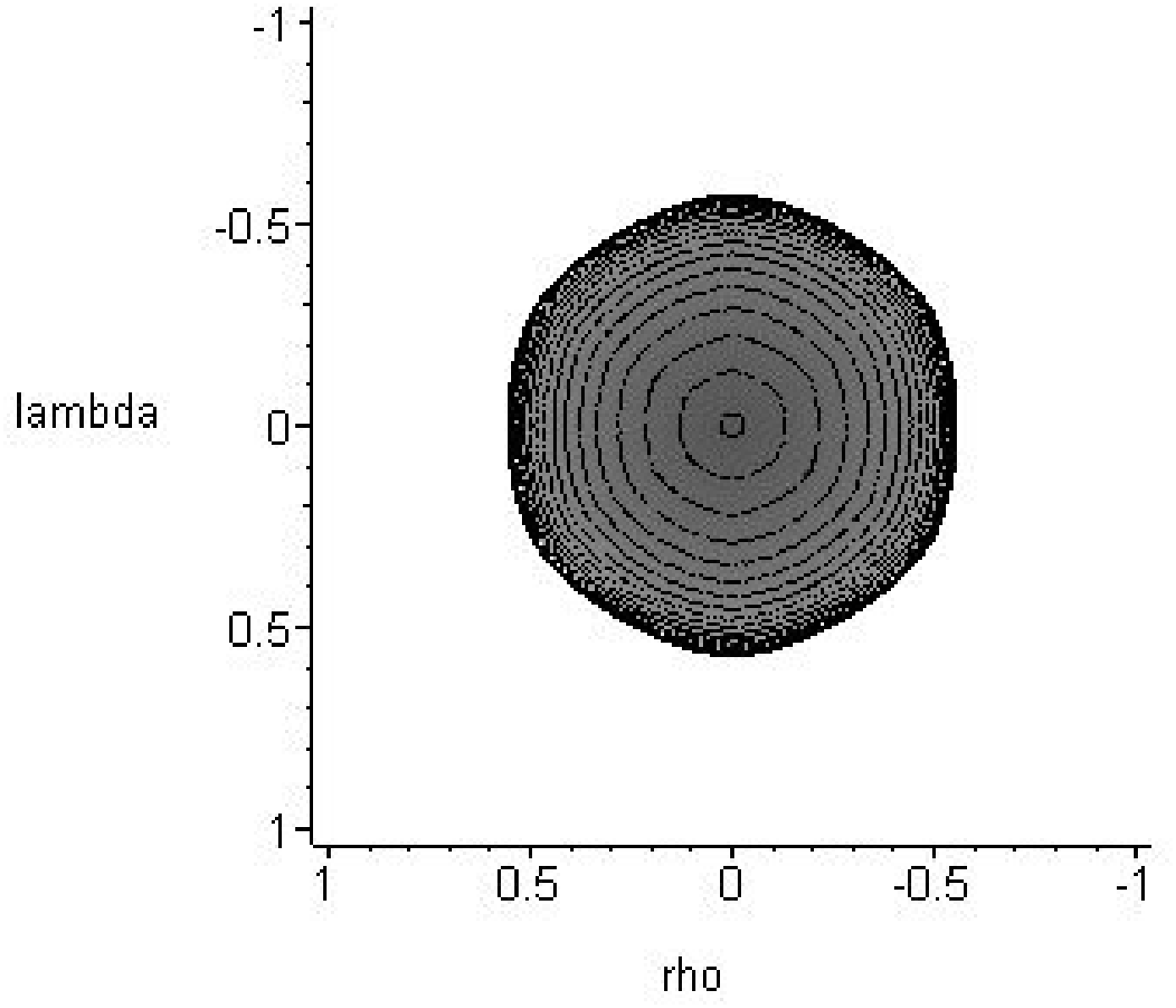';file-properties "XNPEU";}} \\ \hline
\end{tabular}%
\caption{Potential plot for $\Lambda=0$ from both side and top views with solid
lines denoting equipotentials.\label{key}}%
%TCIMACRO{\TeXButton{E}{\end{table}}}%
%BeginExpansion
\end{table}%
%EndExpansion

%TCIMACRO{\TeXButton{B}{\begin{table}[tbp] \centering}}%
%BeginExpansion
\begin{table}[tbp] \centering%
%EndExpansion
\begin{tabular}{|ll|}
\hline
\FRAME{itbpFU}{2.0081in}{2.0081in}{0in}{\Qcb{$\Lambda =0$ \ \ \ \ $p_{%
\protect\theta }=0.5$}}{}{zl_5pt_side.eps}{\special{language "Scientific
Word";type "GRAPHIC";maintain-aspect-ratio TRUE;display "USEDEF";valid_file
"F";width 2.0081in;height 2.0081in;depth 0in;original-width
7.9442in;original-height 7.9442in;cropleft "0";croptop "1";cropright
"1";cropbottom "0";filename '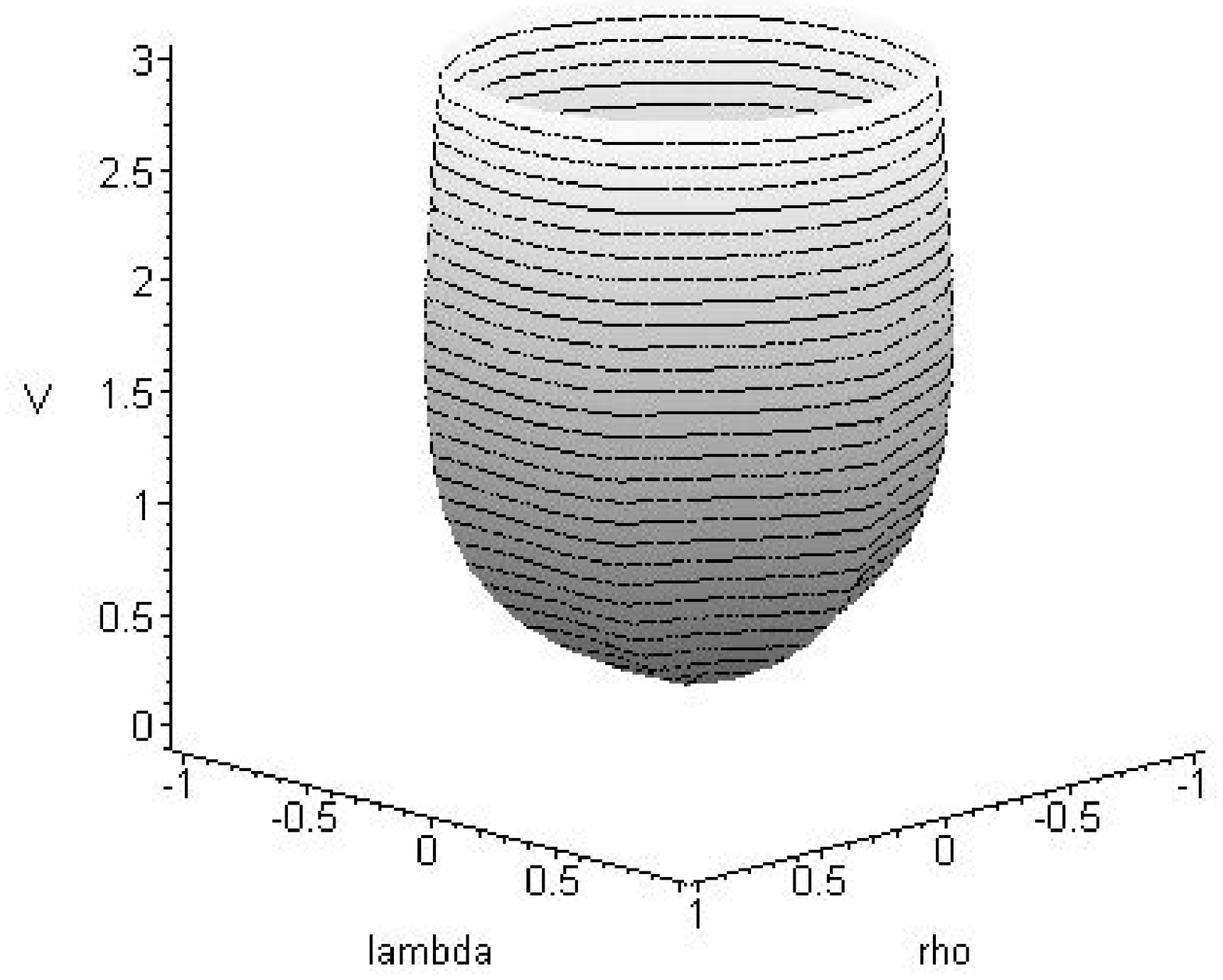';file-properties "XNPEU";}} & 
\FRAME{itbpF}{2.0081in}{2.0081in}{0in}{}{}{zl_5pt_top.eps}{\special{language
"Scientific Word";type "GRAPHIC";maintain-aspect-ratio TRUE;display
"USEDEF";valid_file "F";width 2.0081in;height 2.0081in;depth
0in;original-width 7.9442in;original-height 7.9442in;cropleft "0";croptop
"1";cropright "1";cropbottom "0";filename '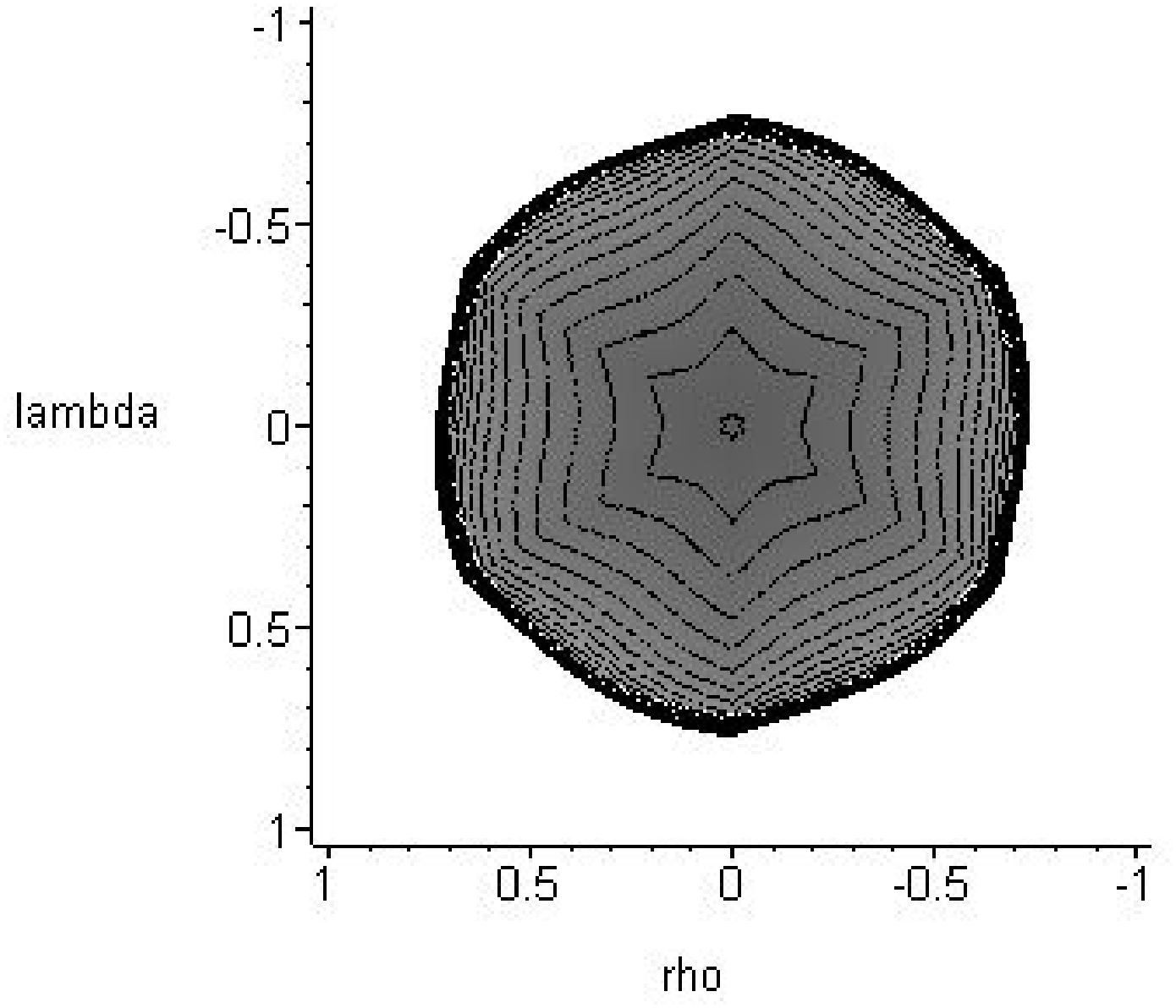';file-properties
"XNPEU";}} \\ \hline
\end{tabular}%
\caption{Potential plot with positive angular momentum from both side and top views with solid
lines denoting equipotentials.\label{key}}%
%TCIMACRO{\TeXButton{E}{\end{table}}}%
%BeginExpansion
\end{table}%
%EndExpansion

%TCIMACRO{\TeXButton{B}{\begin{table}[tbp] \centering}}%
%BeginExpansion
\begin{table}[tbp] \centering%
%EndExpansion
\begin{tabular}{|ll|}
\hline
\FRAME{itbpFU}{2.0081in}{2.0081in}{0.1877in}{\Qcb{$\Lambda =0.1$ \ \ \ $%
p_{r}=0.3$}}{}{pl_pp_side.eps}{\special{language "Scientific Word";type
"GRAPHIC";maintain-aspect-ratio TRUE;display "USEDEF";valid_file "F";width
2.0081in;height 2.0081in;depth 0.1877in;original-width
7.9442in;original-height 7.9442in;cropleft "0";croptop "1";cropright
"1";cropbottom "0";filename '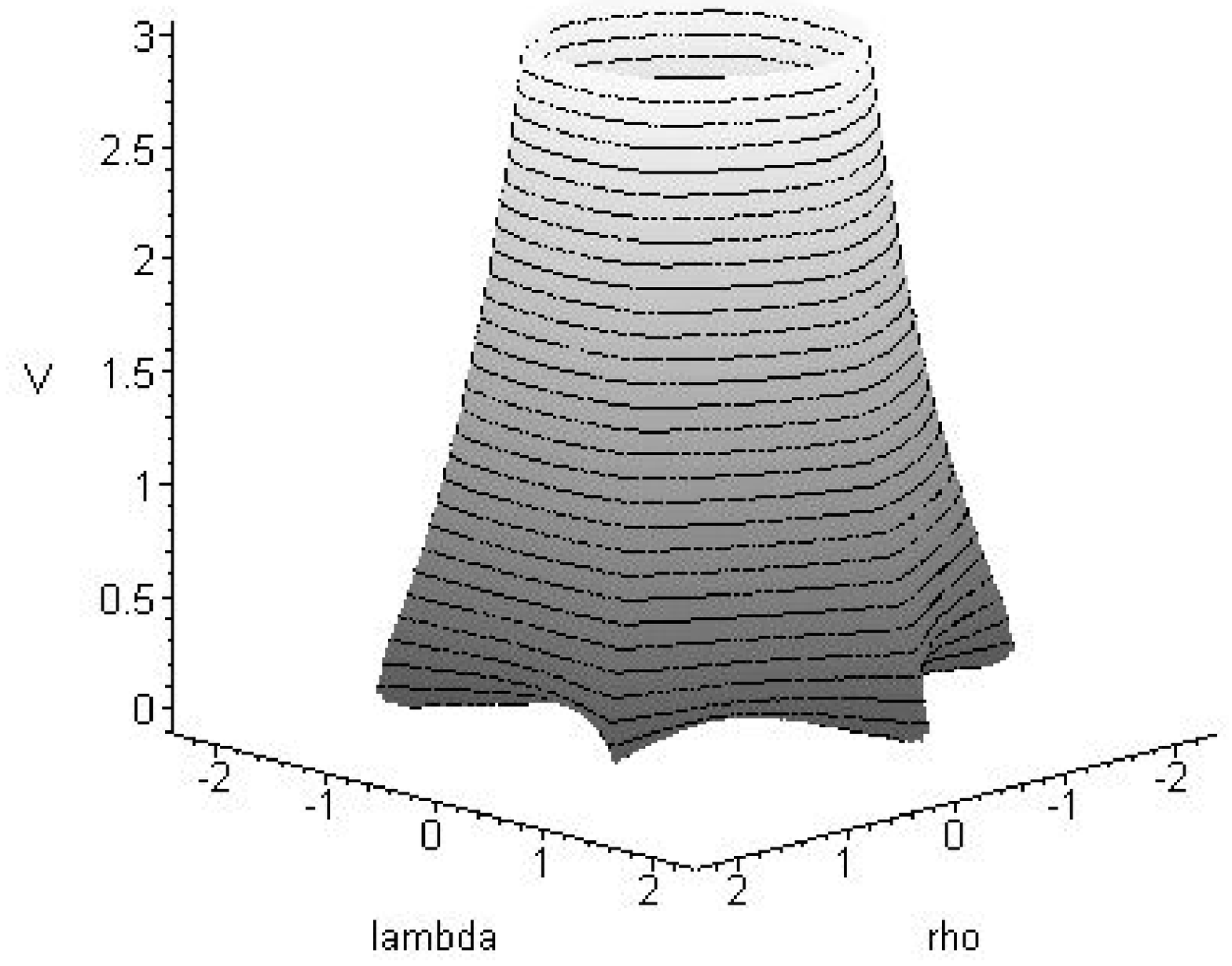';file-properties "XNPEU";}} & 
\FRAME{itbpF}{2.0081in}{2.0081in}{0in}{}{}{pl_pp_top.eps}{\special{language
"Scientific Word";type "GRAPHIC";maintain-aspect-ratio TRUE;display
"USEDEF";valid_file "F";width 2.0081in;height 2.0081in;depth
0in;original-width 7.9442in;original-height 7.9442in;cropleft "0";croptop
"1";cropright "1";cropbottom "0";filename '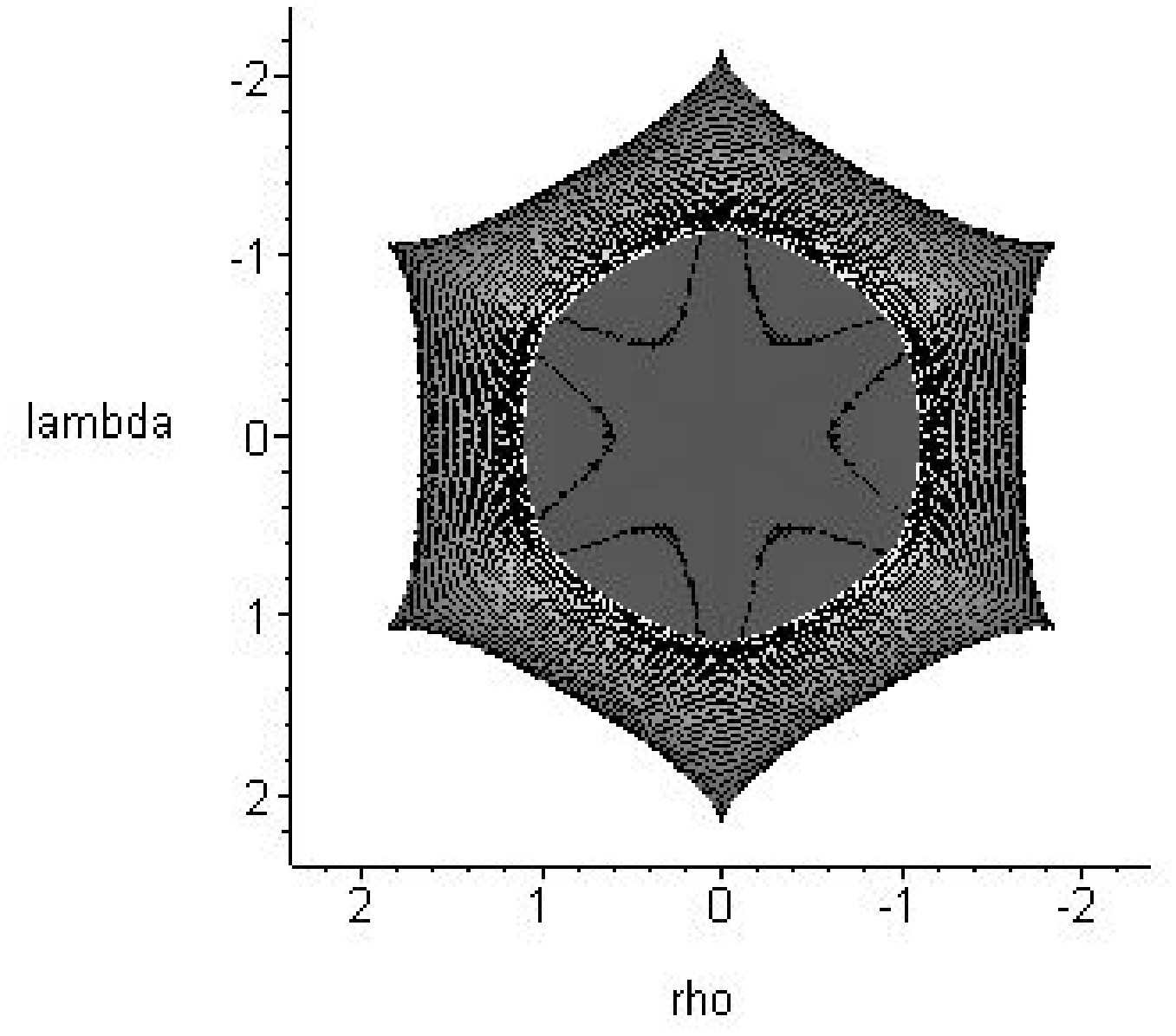';file-properties
"XNPEU";}} \\ \hline
\FRAME{itbpFU}{2.0081in}{2.0081in}{0in}{\Qcb{$\Lambda =0.1$ \ \ \ $%
p_{r}=-0.3 $}}{}{pl_np_side.eps}{\special{language "Scientific Word";type
"GRAPHIC";maintain-aspect-ratio TRUE;display "USEDEF";valid_file "F";width
2.0081in;height 2.0081in;depth 0in;original-width 7.9442in;original-height
7.9442in;cropleft "0";croptop "1";cropright "1";cropbottom "0";filename
'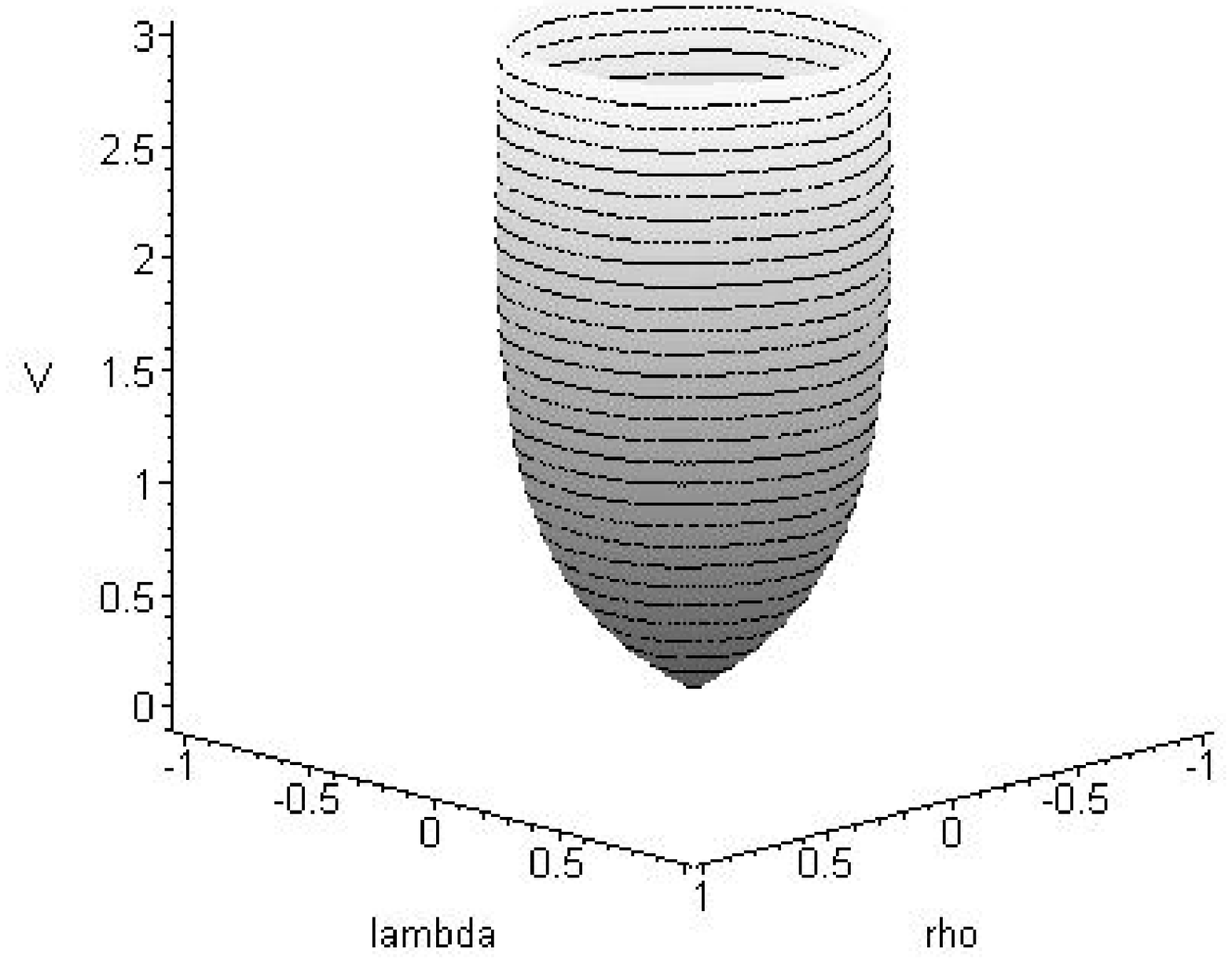';file-properties "XNPEU";}} & \FRAME{itbpF}{2.0081in}{%
2.0081in}{0in}{}{}{pl_np_top.eps}{\special{language "Scientific Word";type
"GRAPHIC";maintain-aspect-ratio TRUE;display "USEDEF";valid_file "F";width
2.0081in;height 2.0081in;depth 0in;original-width 7.9442in;original-height
7.9442in;cropleft "0";croptop "1";cropright "1";cropbottom "0";filename
'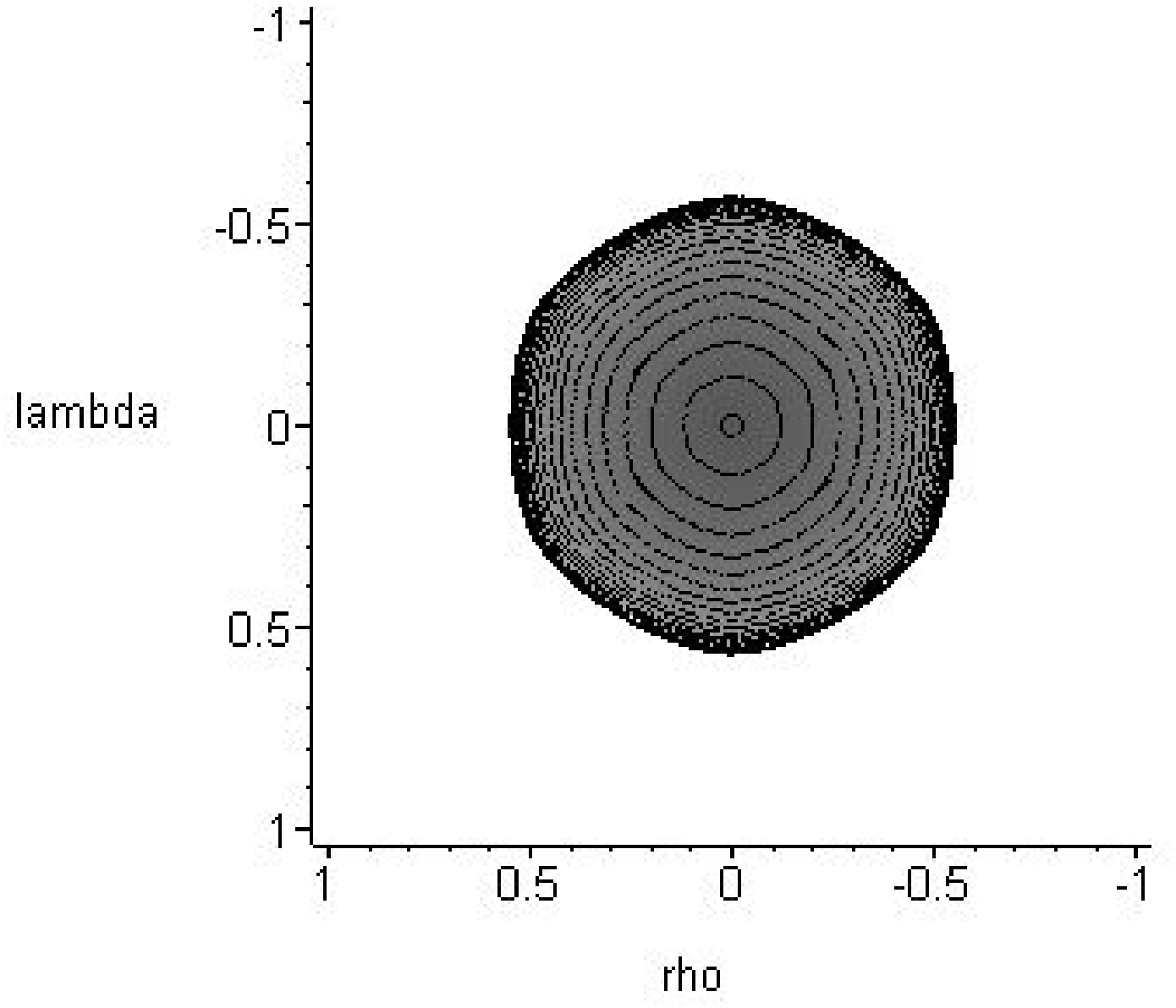';file-properties "XNPEU";}} \\ \hline
\end{tabular}%
\caption{Potential plots for $\Lambda > 0$ from both side and top views with solid
lines denoting equipotentials. \label{key}}%
%TCIMACRO{\TeXButton{E}{\end{table}}}%
%BeginExpansion
\end{table}%
%EndExpansion

%TCIMACRO{\TeXButton{B}{\begin{table}[tbp] \centering}}%
%BeginExpansion
\begin{table}[tbp] \centering%
%EndExpansion
\begin{tabular}{|ll|}
\hline
\FRAME{itbpFU}{2.0081in}{2.0081in}{0in}{\Qcb{$\Lambda =-0.1$ \ \ \ $%
p_{r}=0.3 $}}{}{nl_pp_side.eps}{\special{language "Scientific Word";type
"GRAPHIC";maintain-aspect-ratio TRUE;display "USEDEF";valid_file "F";width
2.0081in;height 2.0081in;depth 0in;original-width 7.9442in;original-height
7.9442in;cropleft "0";croptop "1";cropright "1";cropbottom "0";filename
'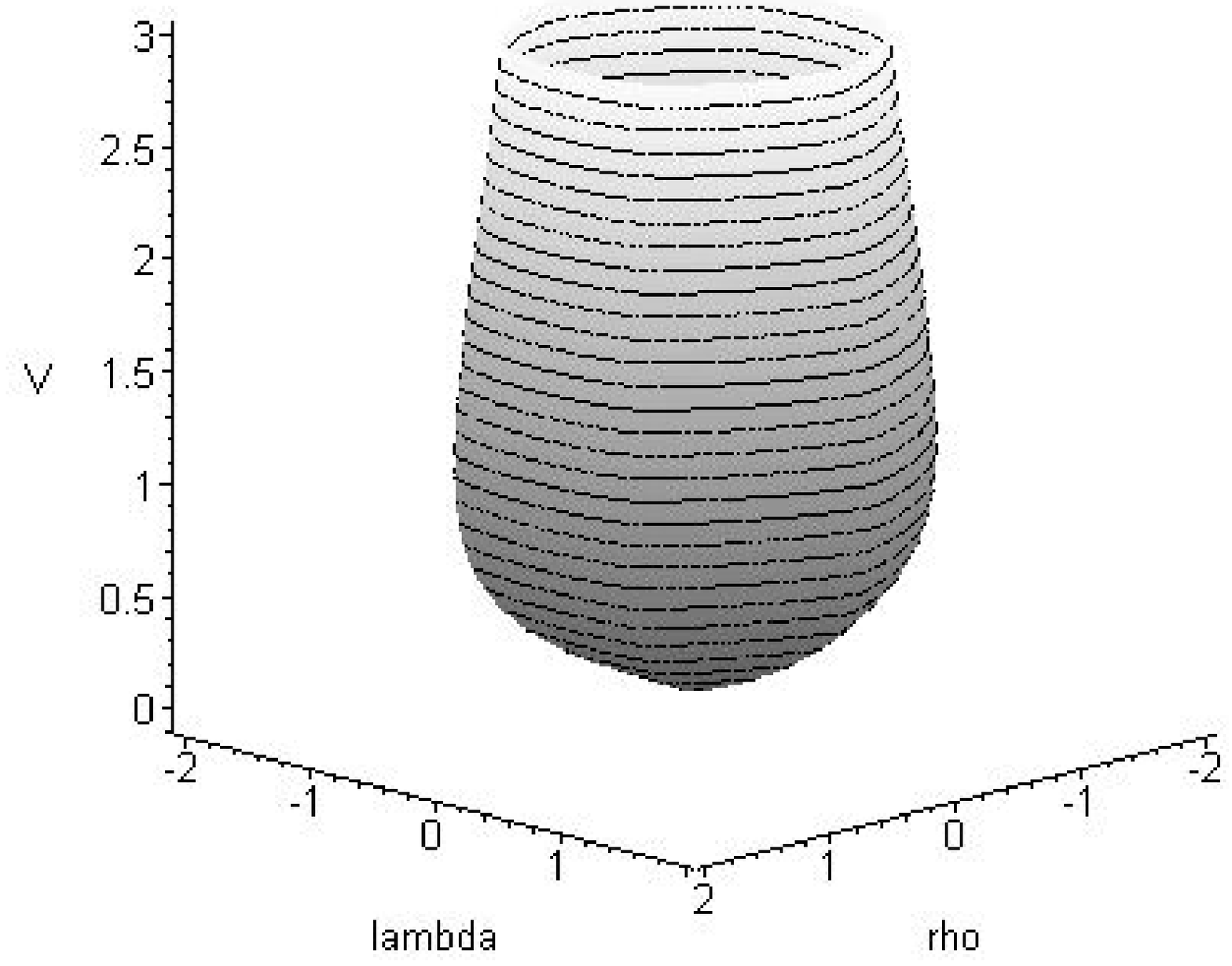';file-properties "XNPEU";}} & \FRAME{itbpF}{2.0358in}{%
2.0358in}{0in}{}{}{nl_pp_top.eps}{\special{language "Scientific Word";type
"GRAPHIC";maintain-aspect-ratio TRUE;display "USEDEF";valid_file "F";width
2.0358in;height 2.0358in;depth 0in;original-width 7.9442in;original-height
7.9442in;cropleft "0";croptop "1";cropright "1";cropbottom "0";filename
'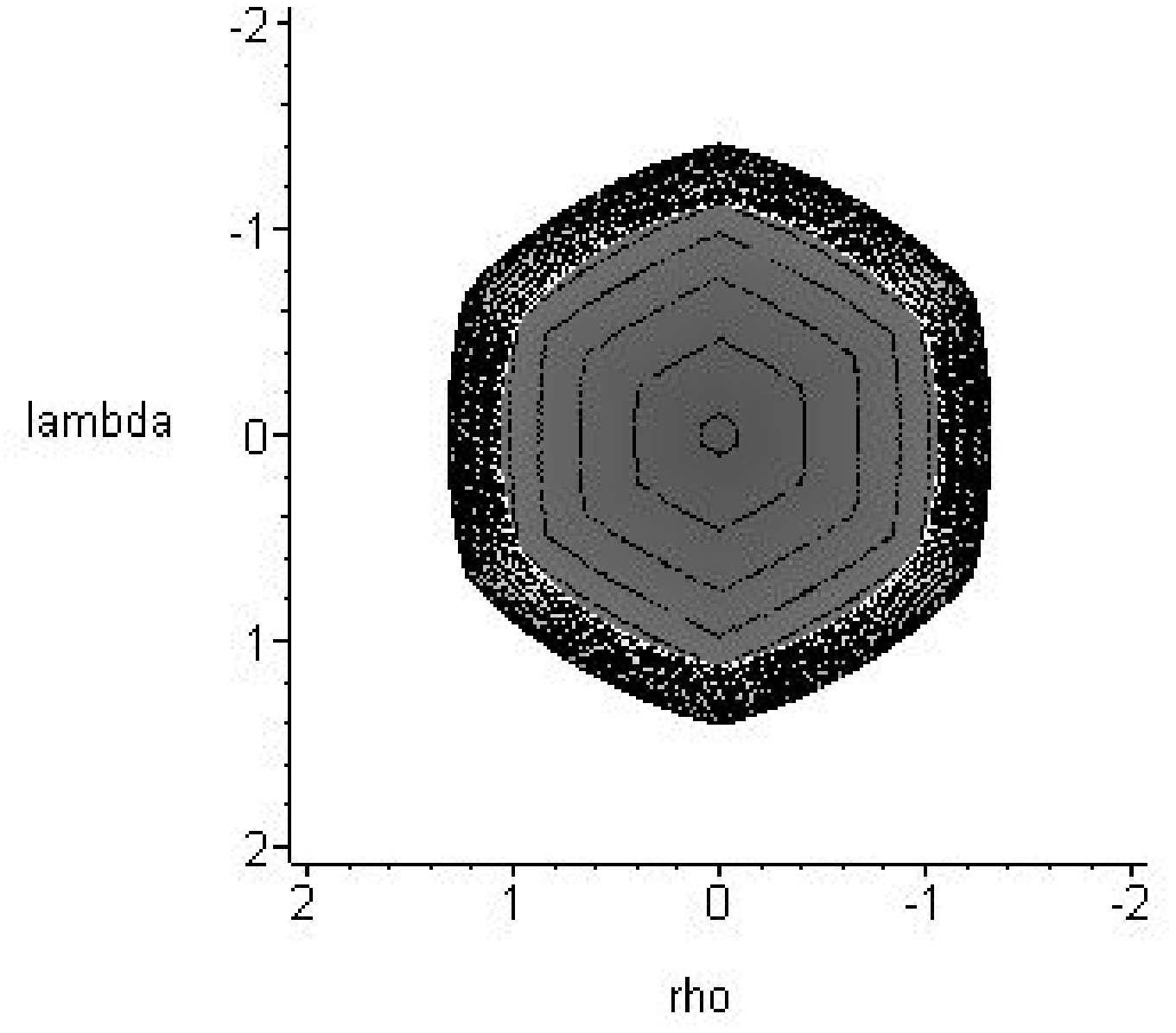';file-properties "XNPEU";}} \\ \hline
\FRAME{itbpFU}{2.0081in}{2.0081in}{0in}{\Qcb{$\Lambda =-0.1$ \ \ \ $%
p_{r}=-0.3$}}{}{nl_np_side.eps}{\special{language "Scientific Word";type
"GRAPHIC";maintain-aspect-ratio TRUE;display "USEDEF";valid_file "F";width
2.0081in;height 2.0081in;depth 0in;original-width 7.9442in;original-height
7.9442in;cropleft "0";croptop "1";cropright "1";cropbottom "0";filename
'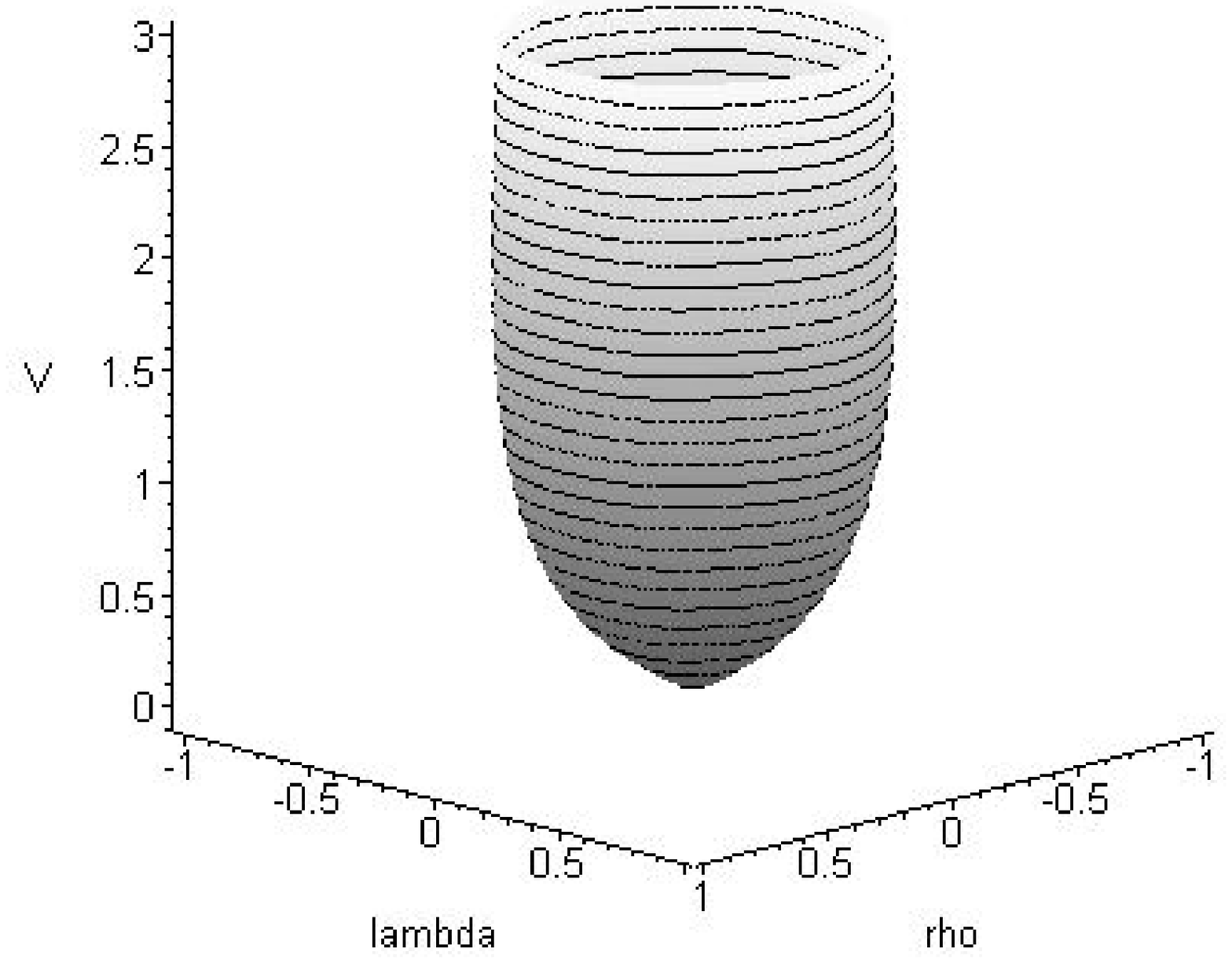';file-properties "XNPEU";}} & \FRAME{itbpF}{2.0081in}{%
2.0081in}{0in}{}{}{nl_np_top.eps}{\special{language "Scientific Word";type
"GRAPHIC";maintain-aspect-ratio TRUE;display "USEDEF";valid_file "F";width
2.0081in;height 2.0081in;depth 0in;original-width 7.9442in;original-height
7.9442in;cropleft "0";croptop "1";cropright "1";cropbottom "0";filename
'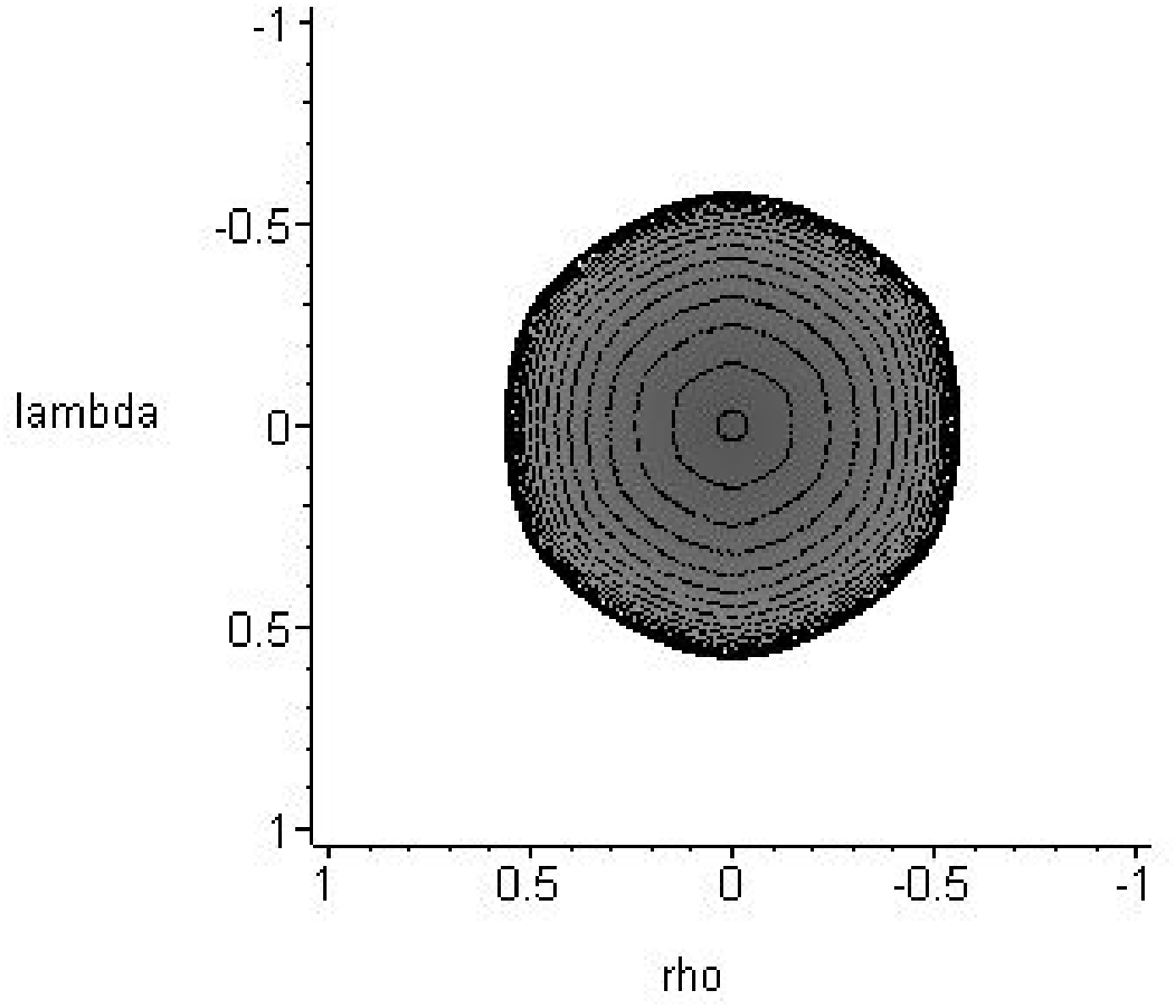';file-properties "XNPEU";}} \\ \hline
\end{tabular}%
\caption{Potential plots for  $\Lambda<0$ from both side and top views with solid
lines denoting equipotentials. \label{key}}%
%TCIMACRO{\TeXButton{E}{\end{table}}}%
%BeginExpansion
\end{table}%
%EndExpansion

First, when $\Lambda =0$ \ and the hex particle has positive radial momentum 
$p_{r}$, the width of the potential at lower energies increases with
increasing $p_{r}$\ and the sides of the hexagonal cross section become more
convex at lower energies, resulting in a star shaped cross section, as shown
in table (1). This widens the bottom of the potential, and lowers the value
of $V_{Rc}$\ at which the cross-section of $V$\ is at its largest. When the
hex-particle has negative $p_{r}$\ the opposite happens: the width of the
potential decreases and the sides of the hexagonal cross section become more
concave, resulting in a more rounded cross section. Though its sides are
initially much steeper, the potential curves back in on itself much more
gradually resulting in a higher value of $V_{Rc}$.

A similar analysis for nonzero angular momentum $p_{\theta }$, as shown in
table (2), indicates that only for very high values does it have any
substantive effect on the potential . As $p_{\theta }$\ increases the sides
of the hexagonal cross section become concave and slightly skewed in the
direction opposing the angular momentum. We expect that this is the cause of
the previously observed rotation of the annulus orbits at higher energies %
\cite{burnell} for the $\Lambda =0$\ case.

When $\Lambda >0$\ the potential, shown in table (3), shifts down slightly
when $p_{r}$ is positive. At lower energies the sides are pushed farther out
while becoming more convex in a manner similar to that which occurs for
positive radial momentum in the original ($\Lambda =0$) potential.\ A side
effect of this is that it seems to lower $V_{Rc}$, which may cause the
system to become unstable at lower energies. Associated with this is the
appearance of a positive critical value of the cosmological constant, which
we will discuss in more detail later. For negative $p_{r}$ and $\Lambda >0$%
,\ at lower energies the potential is pulled inward and becomes more
concave, again in a manner similar to the negative radial momentum in the $%
\Lambda =0$\ case. However, the magnitude of this effect of $\Lambda $ on $%
p_{r}<0$ is much less than the effect on $p_{r}>0$.\ The effects noted above
for angular momentum also seem to increase when $\Lambda >0$. Overall, the
presence of a positive cosmological constant seems to enhance the effects
that momentum has on the $\Lambda =0$\ potential.

When $\Lambda <0$, as shown in table (4), the opposite effect happens. For $%
p_{r}>0$ the sides are pulled in relative to the effects of momentum in the $%
\Lambda =0$\ potential, reducing its diameter\ as well as making the sides
less convex. For $p_{r}<0$, \ the sides of the potential are very slightly
pushed back out relative to the effects of momentum in the $\Lambda =0$\
potential, increasing its diameter. The effects of angular momentum are also
counteracted when $\Lambda <0$, so overall, the presence of a negative
cosmological constant seems to reduce momentum-dependence in the potential.\
\qquad

We have found both upper and lower limits on the range of allowed values of
the cosmological constant $\Lambda $ that are dependent on the energy in the
system. As described before, in order to solve the equations of motion we
must calculate the derivatives of $X$ from Eq. (\ref{eqn-Ki}) with respect
to each $z_{i}$ and $p_{i}$, and then use Eqs. (\ref{Ham2}) and (\ref%
{eqn-dHtodX}) to convert $X$-derivatives to $H$-derivatives, yielding the
canonical equations of motion. However, both $X$ and $H$ must be real. Thus
from eq. (\ref{Ham2}) we find that for any given value of $H$, $\Lambda $
must be chosen in a range that satisfies 
\[
H\geq \frac{4}{\kappa }\sqrt{-\frac{\Lambda }{2}} 
\]%
resulting in a negative critical value for $\Lambda $%
\begin{equation}
\Lambda _{negcrit}=-\frac{H^{2}\kappa ^{2}}{8}  \label{eqn-negcrit}
\end{equation}%
As $\Lambda \rightarrow \Lambda _{negcrit}^{+}$, we see that $X\rightarrow 0$%
\ . From eq. (\ref{eqn-dHtodX}) all of the derivatives of $H$\ tend to zero,
rendering the particle motionless. If we take a fixed value for $H_{0}$ and
increase it by an arbitrarily small amount $h$, when $\Lambda $ is the
negative critical value for an energy of $H_{0}$, from eqs. (\ref{Ham2}, \ref%
{eqn-negcrit}) we get 
\[
H=H_{0}+h=\frac{4}{\kappa }\sqrt{\kappa ^{2}X^{2}+\frac{\Lambda _{negcrit}}{2%
}}=H_{0}\sqrt{1+\frac{16X^{2}}{H_{0}^{2}}} 
\]%
When $h$\ is very small, so is $X$, and we can take the approximation $%
h\simeq 8\frac{X^{2}}{H_{0}}$\ , yielding 
\begin{equation}
\frac{\partial H}{\partial x_{n}}=\frac{\partial h}{\partial x_{n}}=\frac{16X%
}{H_{0}}\frac{\partial X}{\partial x_{n}}=\sqrt{\frac{32h}{H_{0}}}\frac{%
\partial X}{\partial x_{n}}  \label{eqn-nearcrit}
\end{equation}%
for the equations of motion, where $x_{n}$\ is any of the canonical
variables. For small $h$ the equations of motion are all scaled by $\sqrt{h}$%
, which is equivalent to scaling the time factor by $\frac{1}{\sqrt{h}}$.\
Consequently, when the cosmological constant is very close to its negative
critical value for a system with a given energy, the hex-particle will not
so much be restricted in its movement in the $\left( \rho ,\lambda \right) $
plane, and follow that orbit at a much slower pace.\ However, as we will
describe in more detail later, we have found that for small negative values
for the cosmological constant the frequency of the hex-particle's movement
increases, so this trend only appears for very small $h$.\ 

We have not been able to identify analytically the positive critical value
of $\Lambda $ . Rather we identify it numerically: when running simulations
with $\Lambda $\ too large the three particle system breaks out of its bound
state, and our simulation breaks down. We have not yet been able to
conclusively find why the system fails where it does, but we believe it may
be due to the fact that, when radial momentum is present in the hex-particle
system, a positive cosmological constant seems to lower the value of $V_{Rc}$%
, rendering the system intrinsically nonperturbative where it was previously
in a perturbative regime. This hypothesis is supported by the fact that, for
the energies that we have tested, the maximum value of $\Lambda $ allowed
seems to decrease substantially as $H$ increases.

The effect on the hex-particle's orbits from approaching these positive and
negative critical values of the cosmological constant will be further
discussed in later sections.

\qquad

\section{Methods for Solving the Equations of Motion}

We find it easier to study the equal mass three body system by analyzing the
motion of the hex-particle in the $\left( \rho ,\lambda \right) $ plane.\ In
this system the bisectors joining opposite vertices of the potential well
correspond to two of the three particles crossing one another, causing a
discontinuous change in the hex-particle's acceleration. Therefore, in the
Hamiltonian formalism, the motion of the hex-particle is governed by a pair
of differential equations which are continuous everywhere except at the
three hexagonal bisectors $\rho =0$, $\rho -\sqrt{3}\lambda =0$, and $\rho +%
\sqrt{3}\lambda =0$. These correspond to the crossings of particles 1 and 2,
2 and 3, or 1 and 3 respectively.

We allow the hex-particle to freely cross over each of these bisectors,
which corresponds to a pair of particles passing through each other. An
analogous system in the Newtonian case, consists of the motions of a ball
under a constant gravitational force elastically colliding with a wedge \cite%
{LMiller}. This system is nearly identical to the one we study insofar as,
in the equal mass case, an elastic collision between a pair of particles in
the three body system cannot be distinguished from a crossing of two equal
mass particles. However, a distinction between the two systems in a certain
class of orbits has been previously observed \cite{burnell}.

The nonsmoothness of the potential gives rise to interesting dynamics in the
system.\ We shall refer to two distinct types of motion in our subsequent
analysis \cite{LMiller}.\ $A$ motion, corresponding to the same pair of
particles crossing twice in a row in the three body system (or equivalently
the hex-particle crossing a single bisector twice in succession), and $B$
motion, in which a single particle crosses both of the other particles in
succession (or equivalently the hex-particle crossing two successive
hexagonal bisectors).\ For a given orbit we can characterize the motion by a
sequence of these symbols with a finite exponent $n$ denoting $n$ repeats
and an overbar denoting an infinite repeated sequence. It is important to
note that the classification of a crossing as $A$ or $B$ motion is dependent
on the previous crossing, so there is an ambiguity in the classification of
either the final or the initial crossing. This is resolved by taking the
initial crossing as unlabeled; since we are considering large sequences of
motion this ambiguity causes no practical difficulties \cite{burnell}.

There is a technical difficulty in comparing trajectories in systems with
different values for the cosmological constant. Such changes induce a change
in the determining equation and so we cannot use exactly the same initial
conditions. We will deal with this by comparing trajectories with the same
fixed-energy (FE) initial conditions. This is done by fixing the initial
values of $H$, $\rho $, $\lambda $\ and $p_{\rho }$, and adjusting $%
p_{\lambda }$\ so that the Hamiltonian constraint is satisfied.

We have implemented three methods of analysis to study the motion of the
system. First, we plot the trajectory of the hex-particle in the $\left(
\rho ,\lambda \right) $ plane and compare the motion under different initial
conditions and in systems with a different value for the cosmological
constant.\ Second, we plot the motion of the three particles as a function
of time which gives us a different method for visualization, allowing us to
identify differences in the various types of motion. Third, we construct
Poincare sections by plotting the radial momentum ($p_{r}$, labeled as $x$)
against the square of the angular momentum ($p_{\theta }^{2}$, labeled as $z$%
) of the hex-particle each time it crosses one of the bisectors. When all
particles have equal mass, all of the bisectors are equivalent so all of the
crossings can be plotted on the same surface of section. This allows us to
easily identify regions of periodicity, quasiperiodicity, and chaos, and we
will discuss each of these features.

There is no closed-form solution for either the determining equation (\ref%
{det-eqn1}) or to the equations of motion so we must solve these equations
numerically. We proceed by introducing dimensionless variables 
\[
X=M_{tot}\hat{x}\quad H=M_{tot}\hat{h}\quad \Lambda =M_{tot}\kappa \hat{%
\Lambda} 
\]%
where $M_{tot}=\sum_{a=1}^{3}m_{a}$, and also%
\[
p_{i}=M_{tot}\hat{p}_{i}\quad z_{i}=\frac{4}{\kappa M_{tot}}\hat{z}_{i} 
\]%
Substituting these variables into Eqs. (\ref{eqn-Rpm}), (\ref{eqn-Mij}), (%
\ref{eqn-Li}), (\ref{eqn-Lstar}) we get%
\[
R_{i\pm }=\kappa M_{tot}\sqrt{\left[ \hat{x}-\frac{\epsilon }{4}\left(
\sum_{a=1}^{3}\hat{p}_{a}s_{ai}\pm \hat{p}_{i}\right) \right] ^{2}-\frac{%
\hat{\Lambda}}{2}}\equiv \kappa M_{tot}\hat{R}_{i\pm } 
\]%
\[
\sqrt{p_{a}^{2}+m_{a}^{2}}=M_{tot}\sqrt{\hat{p}_{a}^{2}+\hat{m}_{a}^{2}} 
\]%
which gives us%
\[
M_{ij}\equiv \kappa M_{tot}\hat{M}_{ij}\quad L_{i}\equiv \kappa M_{tot}\hat{L%
}_{i}\quad L_{i}^{\ast }\equiv \kappa M_{tot}\hat{L}_{i}^{\ast } 
\]%
The notation for these new variables have exactly the same form but without
the factor of kappa inside the square roots and all multiplied by a factor
of $\kappa M_{tot}$. Thus when writing the determining equation in
dimensionless variables the factors of $\left( \kappa M_{tot}\right) ^{3}$
cancel out everywhere and the exponential factors become

{\it 
\[
s_{ji}\left[ \kappa M_{tot}\left( \hat{M}_{ij}+\hat{L}_{i}\right) \frac{1}{%
\kappa M_{tot}}\hat{z}_{ki}-\kappa M_{tot}\left( \hat{M}_{ji}+\hat{L}%
_{j}\right) \frac{1}{\kappa M_{tot}}\hat{z}_{kj}\right] =s_{ji}\left[ \left( 
\hat{M}_{ij}+\hat{L}_{i}\right) \hat{z}_{ki}-\left( \hat{M}_{ji}+\hat{L}%
_{j}\right) \hat{z}_{kj}\right] 
\]%
}

So we see that introducing dimensionless variables is equivalent to writing
the determining equation with $\kappa M_{tot}$\ set equal to unity.

We numerically integrate the equations\ of motion generated from the
determining equation (\ref{det-eqn3}) using a MatLab ODE routine (ODE45 or
ODE113). When the particles are not in the arrangement $z_{1}<z_{2}<z_{3}$\
our program temporarily changes the labels on the particles so they satisfy
this condition. The equations of motion are then calculated for each of the
particles and their original labels are returned. This method works
regardless of whether or not the particles are identical.\ 

For the Poincare sections we stopped the integration each time the
hex-particle crossed one of the bisectors by using an ``events'' function
and saving the values of radial and angular momentum for plotting.\ Ideally
for each chaotic trajectory the Poincare section should be allowed to run
for a long time in order to accurately determine which regions of the plane
the chaotic motion extended to and which regions were restricted.

In order to control errors, we imposed absolute and relative error
tolerances of $10^{-8}$ for the numerical ODE solvers.\ For the values of H
we studied this yielded numerically stable solutions. We tested this by
checking that the energy in the system remained constant throughout the
motion to a value no larger than $10^{-6}$.

\bigskip

\section{Equal Mass Trajectories}

We look at the cases where all three particles have equal mass and the
hexagonal symmetry of the potential well is maintained.\ As described in
previous papers \cite{burnell} we find three general classes of orbits which
we denote as annuli, pretzel, and chaotic.\ Examples of these classes of
orbits have been found at all energies and in the presence of a positive and
negative cosmological constant.

Furthermore, in each class there are orbits that eventually densely cover
the region of the $\left( \rho ,\lambda \right) $ space that they occupy and
orbits that do not.\ The latter situation corresponds to a regular orbit
that repeats itself after a finite time -- the symbol sequence consists of a
finite sequence repeated infinitely many times, resulting in a periodic
trajectory. The former situation corresponds to an orbit where the same
finite sequence is repeated, but with A motion added or removed at irregular
intervals, resulting in a quasiperiodic trajectory.

\FRAME{ftbpFU}{4.0395in}{3.2846in}{0pt}{\Qcb{Examples of near-periodic
orbits (run for 200 time units), trajectories do not cover the entire $%
\left( \protect\rho ,\protect\lambda \right) $ space. \ Periodic annulus and
pretzel orbits can be found for almost all testable energies and values of $%
\Lambda $.}}{\Qlb{periodic}}{periodic.eps}{\special{language "Scientific
Word";type "GRAPHIC";maintain-aspect-ratio TRUE;display "USEDEF";valid_file
"F";width 4.0395in;height 3.2846in;depth 0pt;original-width
6.8346in;original-height 5.5486in;cropleft "0";croptop "1";cropright
"1";cropbottom "0";filename 'periodic.eps';file-properties "XNPEU";}}

\FRAME{ftbpFU}{4.8179in}{3.915in}{0pt}{\Qcb{Arbitrary examples of
quasi-periodic orbits that densely fill the $\left( \protect\rho ,\protect%
\lambda \right) $ space (run for 200 time steps). Quasi-periodic annulus and
pretzel orbits can be found at all testable energies and values of $\Lambda $%
.}}{\Qlb{quasiperiodic}}{quasiperiodic.eps}{\special{language "Scientific
Word";type "GRAPHIC";maintain-aspect-ratio TRUE;display "USEDEF";valid_file
"F";width 4.8179in;height 3.915in;depth 0pt;original-width
6.7066in;original-height 5.5486in;cropleft "0";croptop "0.9936";cropright
"1.0132";cropbottom "0";filename '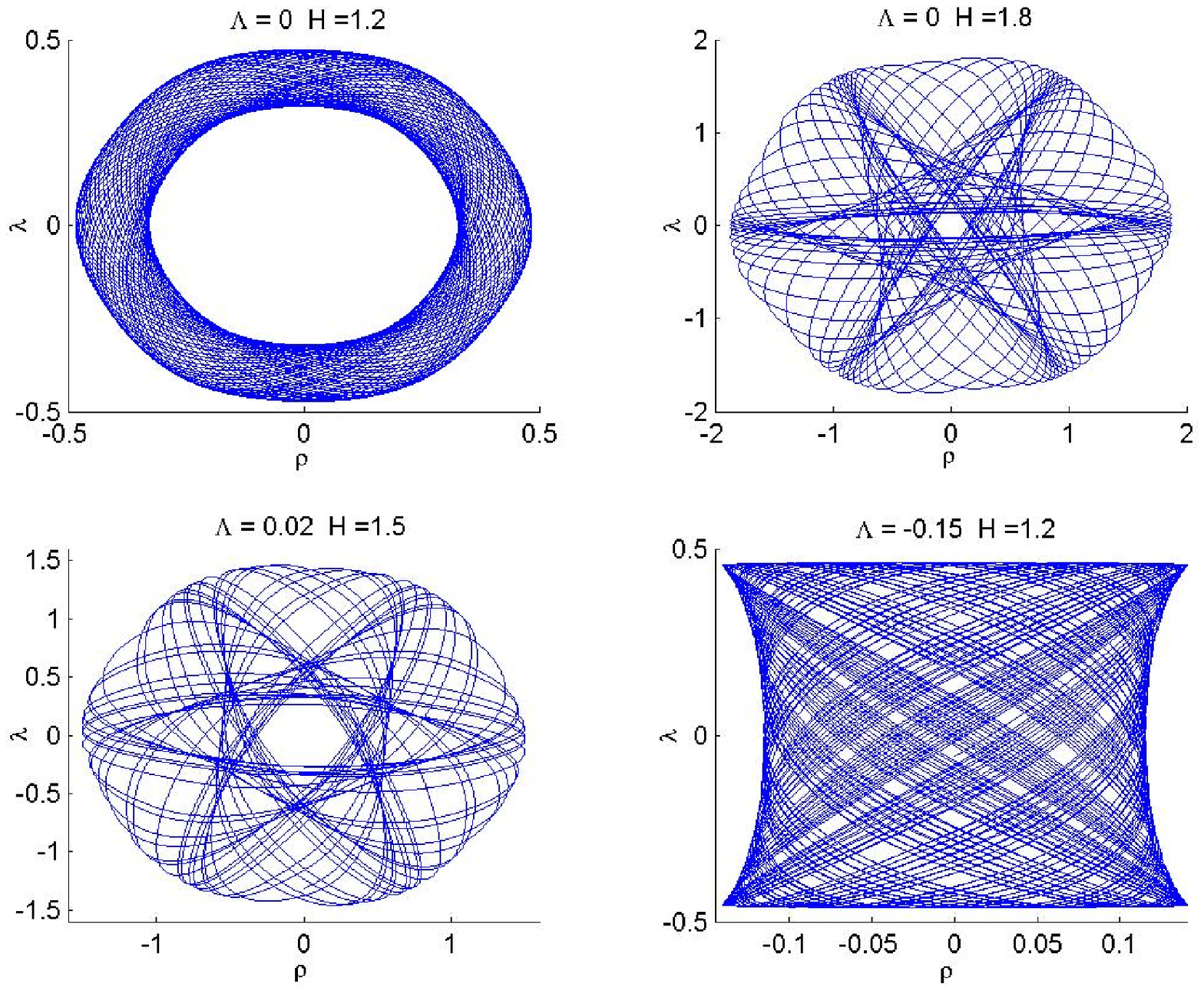';file-properties
"XNPEU";}}

The quasiperiodic orbits closely resemble the periodic orbits, except that
they fail to exactly repeat themselves and have some form of precession that
causes the orbits to fill a region of space.\ Thus these quasiperiodic
orbits show a high degree of regularity, manifest by its periodic symbol
sequence.

\bigskip

\subsection{Annulus Orbits}

The annuli are orbits where the hex-particle never crosses the same bisector
twice in succession, resulting in the symbol sequence $\overline{B}$ and an
orbit in the shape of an annulus encircling the origin in the $\left( \rho
,\lambda \right) $ plane.

These annulus orbits can be both quasiperiodic and fill in the entire ring,
while a few repeat themselves after some number of rotations about the
origin.\ Since periodic orbits are very difficult to find numerically, the
orbits in the figures are actually just very close to being periodic, so
that the pattern of the periodic orbit can be seen.

\FRAME{ftbpFU}{5.0029in}{3.9375in}{0pt}{\Qcb{Annulus orbits for zero
cosmological constant ($H=1.2$ lower left; $H=1.8$ upper right) shown in
conjunction with their corresponding three-particle trajectories.\ The
vertical axis is the displacement from the origin in units of $\protect%
\kappa M_{tot}c^{2}/4$. These orbits have been run for 200 time steps but
the three-particle trajectory plots have been truncated after 30 time
steps.\ The rotation in the annulus orbits is greater for the higher energy
example.}}{\Qlb{ann_example}}{annulusexamples.eps}{\special{language
"Scientific Word";type "GRAPHIC";maintain-aspect-ratio TRUE;display
"USEDEF";valid_file "F";width 5.0029in;height 3.9375in;depth
0pt;original-width 6.6366in;original-height 5.2157in;cropleft "0";croptop
"1";cropright "1";cropbottom "0";filename '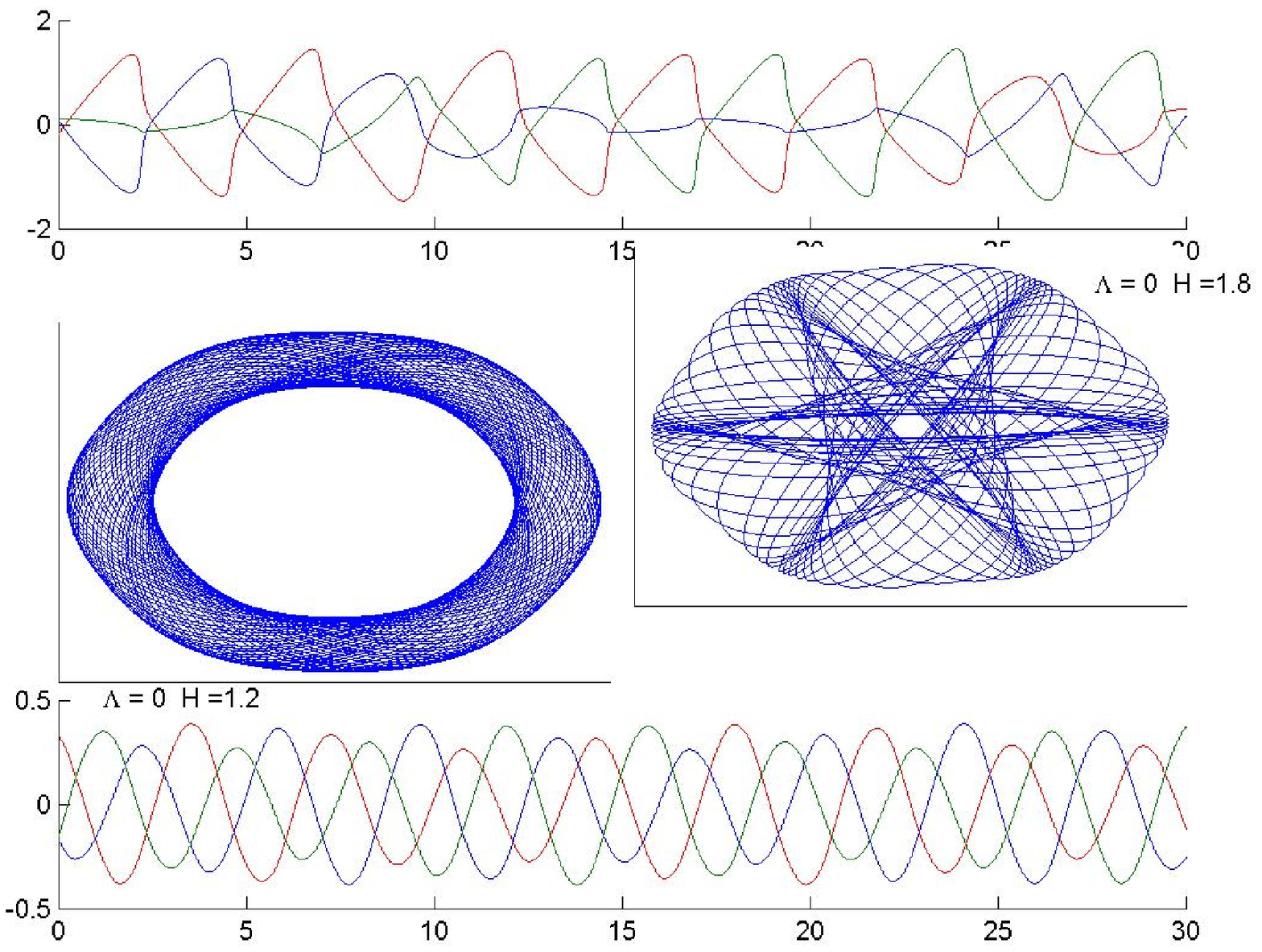';file-properties "XNPEU";}}

There is a slight rotation in the annulus orbits as $H$\ increases, but
otherwise little qualitative difference between the annulus orbits occurs
for different energies. However we do see some qualitative changes to the
orbits as $\Lambda $ changes. For the larger positive values of $\Lambda $
that we can obtain numerically, we find for the same FE initial conditions
that the outer part of the annulus extends farther out into the $\left( \rho
,\lambda \right) $ plane. We also find that the width of the annulus also
increases more than if the orbit were simply linearly dilated (ie
`photographically enlarged'). Conversely, for large negative $\Lambda $ the
orbit covers less of the $\left( \rho ,\lambda \right) $ plane and the
annulus becomes thinner by a greater factor.\ These effects can be seen in
figure (\ref{ann_Lchange}) and seem to be consistent with how the potential
changes with $\Lambda $.\ For positive $\Lambda $, the effect that positive
radial momentum has on pushing out the potential and negative radial
momentum on pulling the potential in towards the origin is\ exaggerated.
Hence when the hex-particle is moving radially outward it will go farther
because the potential is being pushed back. Similarly when the hex-particle
is moving radially inward it will go farther because the potential is being
pulled in and driving it farther.\ For negative $\Lambda $ the opposite
occurs and the radial motion is somewhat damped, resulting in a smaller
range of radial motion.

\FRAME{ftbpFU}{4.951in}{3.9375in}{0pt}{\Qcb{Overlay of annulus orbits at $%
H=1.5$ starting with the same FE initial conditions but with different
values for $\Lambda $ run for 200 time steps and their corresponding
three-particle trajectories truncated after 30 time steps (top $\Lambda
=0.02 $; middle $\Lambda =0$; bottom $\Lambda =-0.25$). \ }}{\Qlb{ann_Lchange%
}}{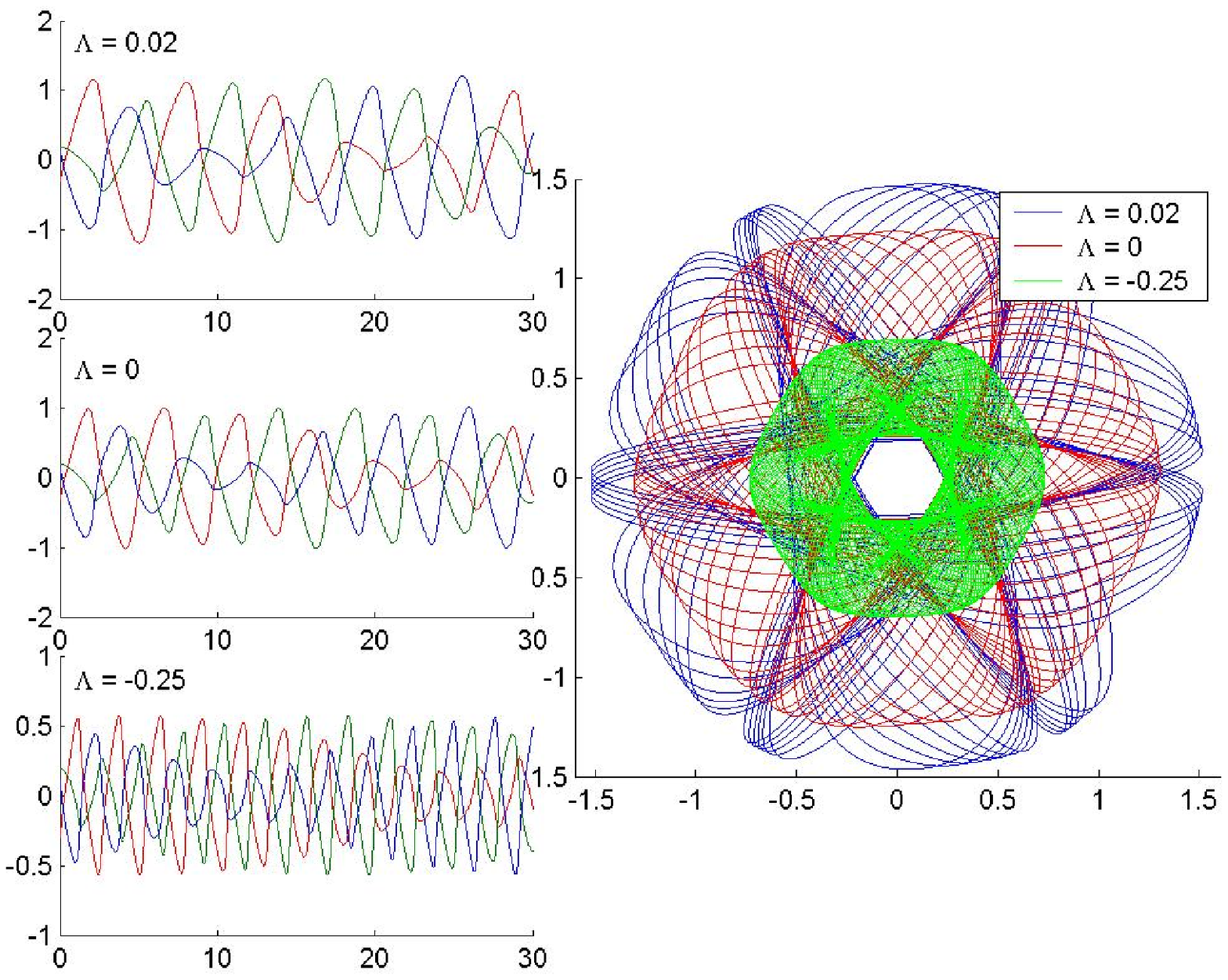}{\special{language "Scientific Word";type
"GRAPHIC";maintain-aspect-ratio TRUE;display "USEDEF";valid_file "F";width
4.951in;height 3.9375in;depth 0pt;original-width 6.5683in;original-height
5.2157in;cropleft "0";croptop "1";cropright "1";cropbottom "0";filename
'AnnulusLambdaChange.eps';file-properties "XNPEU";}}

Another characteristic change induced by $\Lambda $ is that the frequency of
the three particle motion decreases as $\Lambda $ becomes positive and
increases as $\Lambda $ becomes negative. At values of $\Lambda $ extremely
close to the negative critical value given in Eq. (\ref{eqn-negcrit}) the
frequency of the motion should eventually decrease as described in Eq. (\ref%
{eqn-nearcrit}). However, our simulation breaks down due to error tolerance
limits so at this time we have not been able to verify this property
numerically. This non-linear change in frequency can result in the same set
of initial conditions producing periodic or quasiperiodic orbits depending
on $\Lambda $. This change in frequency has been observed in all three types
of orbits.

\bigskip

\subsection{Pretzel Orbits}

In Pretzel orbits, the hex-particle essentially oscillates back and forth
about one of the bisectors, corresponding to a stable or quasistable bound
subsystem of two particles. This bound pair then orbits the remaining
particle, which corresponds to the hex-particle oscillating along the same
bisector. The existence of a two-particle bound subsystem was discovered in
the Newtonian case \cite{Rouet} and has been observed previously in the $%
\Lambda =0$\ case \cite{burnell}. Symbolically, these orbits can be written
as $\prod_{ijk}\left( A^{n_{i}}B^{3m_{j}}\right) ^{l_{k}}$, where $%
n_{i},m_{j},l_{k}\in Z^{+}$, with some $l_{k}$ possibly infinite. This
motion results in an extremely diverse collection of orbits. Many families
of regular orbits exist, containing one base element (for example $AB^{3}$)
and a sequence of elements formed by appending an A to each existing
sequence of A's (for example $\left\{
AB^{3},A^{2}B^{3},A^{3}B^{3},...\right\} $). This results in an extremely
complex structure for the phase space, which we will analyze later.

\FRAME{ftbpFU}{4.9623in}{3.9375in}{0pt}{\Qcb{Examples of pretzel orbits for $%
\Lambda =0$ ($H=1.8$, lower left; $H=1.5$, upper right) run for 200 time
steps and shown in conjunction with their corresponding three-particle
trajectories truncated after 60 time steps. \ Interactions resembling a
quasistable bound subsystem of two particles orbiting a third particle can
clearly be seen.}}{\Qlb{pzl_examples}}{pretzelexamples.eps}{\special%
{language "Scientific Word";type "GRAPHIC";maintain-aspect-ratio
TRUE;display "USEDEF";valid_file "F";width 4.9623in;height 3.9375in;depth
0pt;original-width 6.5838in;original-height 5.2157in;cropleft "0";croptop
"1";cropright "1";cropbottom "0";filename '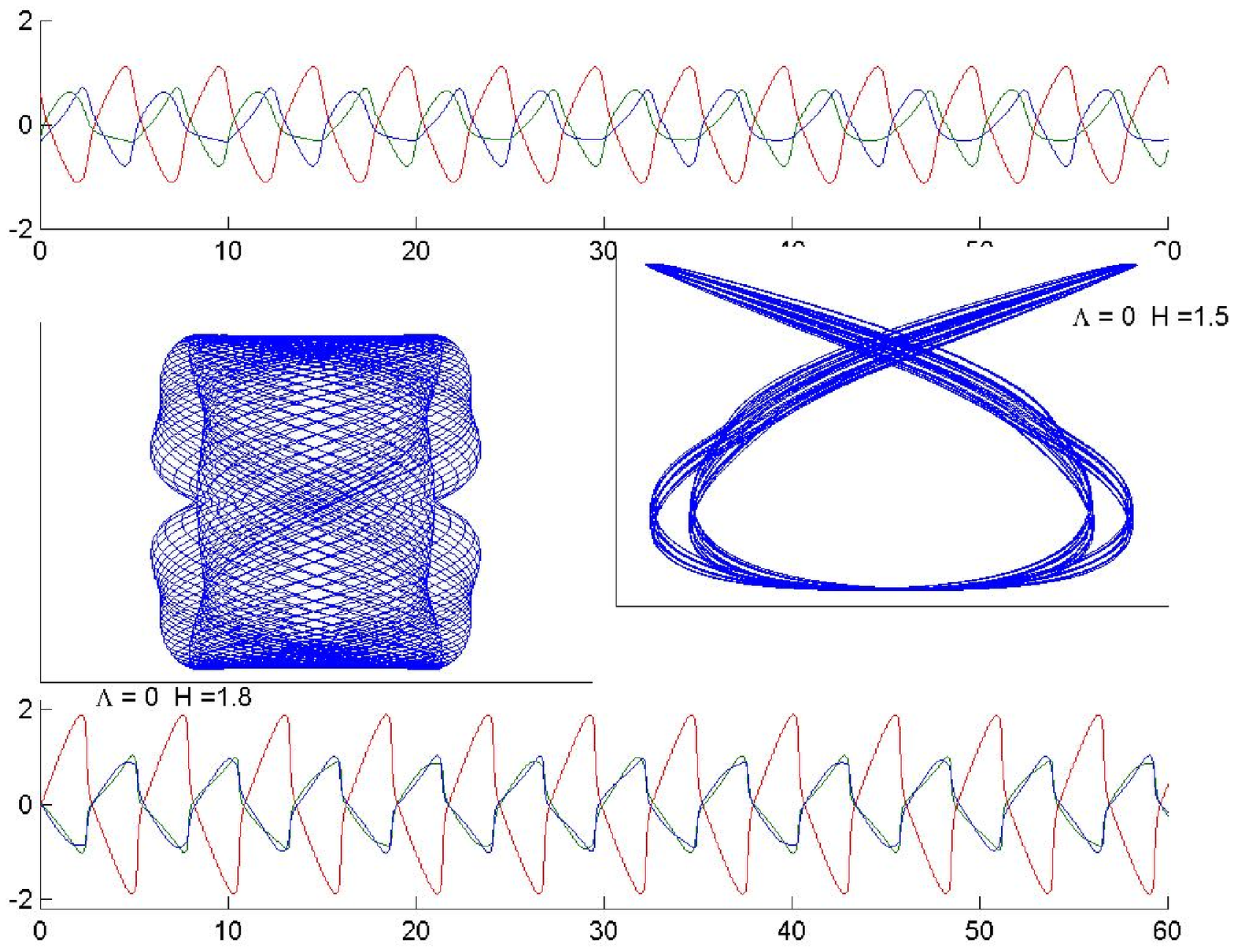';file-properties "XNPEU";}}

The $B^{3}$ term in the above sequences corresponds to a $180^{o}$ swing of
the hex-particle around the origin, and the orbits in the $\left( \rho
,\lambda \right) $ plane that result from this motion comprise a wide
variety of twisted, pretzel-like figures. Again, in this class of orbits we
find both periodic orbits with an infinitely repeating symbol sequence, and
quasiperiodic orbits that densely fill a cylindrical tube in the $\left(
\rho ,\lambda \right) $ plane which usually has a kink about the $\lambda =0$
line.

Qualitatively, the characteristics of the pretzel orbits are very similar at
different energies. At all of the values of $H$ that we can obtain
numerically we can find orbits of all of the different allowed families that
have regular symbol sequences. One of the differences is that it seems that
the kinks in at the $\lambda =0$ line become more pronounced at higher
energies, as previously observed in \cite{burnell}.

\FRAME{ftbpFU}{5.0254in}{3.9375in}{0pt}{\Qcb{Overlay of pretzel orbits
starting with the same FE initial conditions but with different values for $%
\Lambda $ run for 200 time steps and their corresponding three-particle
trajectories truncated after 30 time steps (top $\Lambda =0.02$; middle $%
\Lambda =0$; bottom $\Lambda =-0.25$). \ }}{\Qlb{pzl_Lchange}}{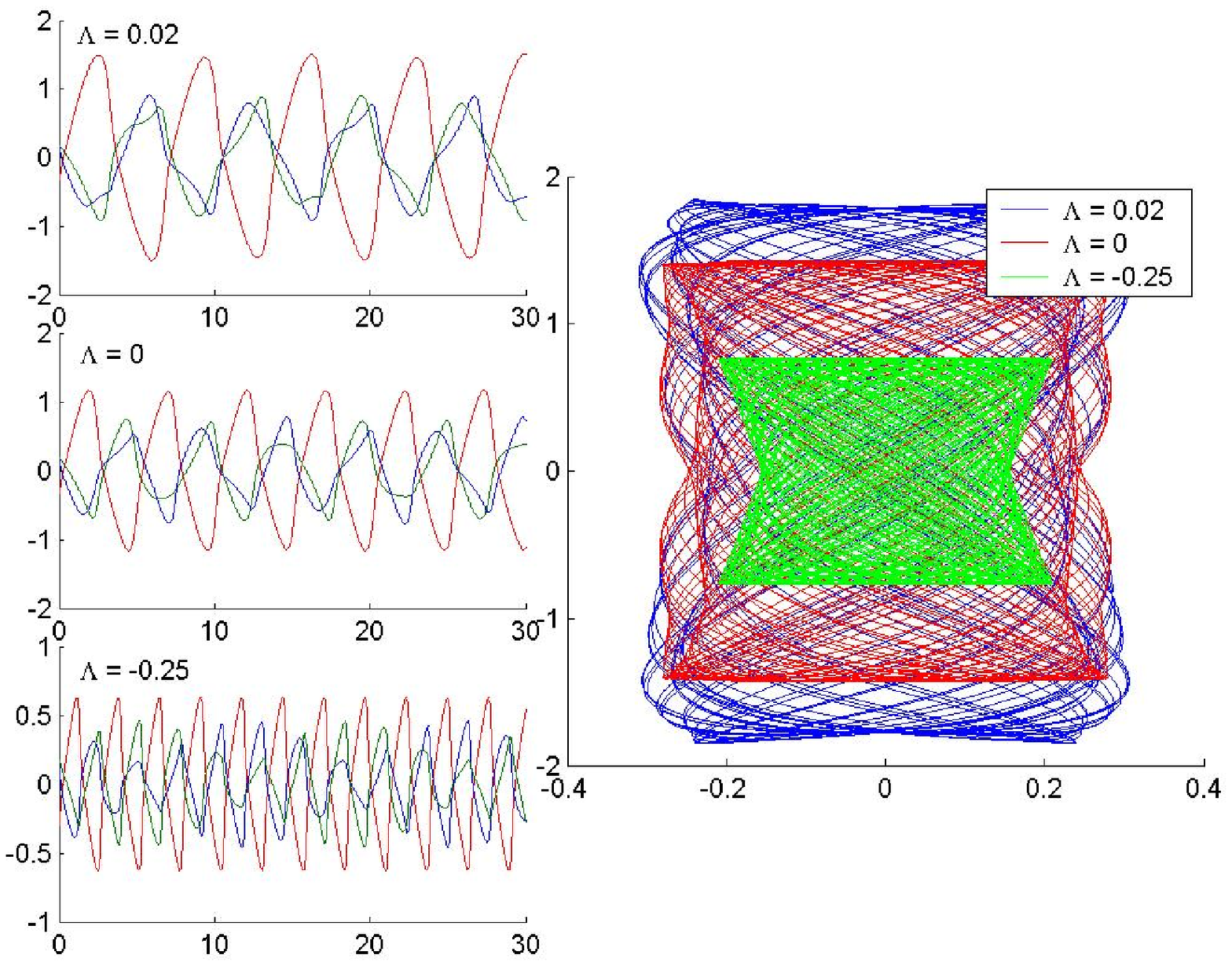}{\special{language "Scientific Word";type
"GRAPHIC";maintain-aspect-ratio TRUE;display "USEDEF";valid_file "F";width
5.0254in;height 3.9375in;depth 0pt;original-width 6.6668in;original-height
5.2157in;cropleft "0";croptop "1";cropright "1";cropbottom "0";filename
'PretzelLambdaChange.eps';file-properties "XNPEU";}}

As we change the cosmological constant we see many similar changes to the
pretzel orbits that we found for the annulus orbits. As $\Lambda $ increases
the orbits extend farther out on the $\left( \rho ,\lambda \right) $ plane,
corresponding to the widening of the potential with positive radial momentum
described before. Alternatively, as $\Lambda $ decreases the orbits cover
less of the $\left( \rho ,\lambda \right) $ plane, corresponding to the
contracting of the potential with negative radial momentum. We can find
orbits with the same regularly repeating symbol sequence for all values of $%
\Lambda $ that we can obtain numerically.\ However, as $\Lambda $\ becomes
positive we have found that some pretzel orbits begin to show a form of
chaotic behaviour not seen in the $\Lambda \leq 0$\ cases. \ We will
describe these cases in more detail later.

Again in the three body figures we find that the frequency of the motion is
increased and reduced for negative and positive $\Lambda $ respectively.\
This non-linear change in frequency can cause the same set of initial
conditions to result in either periodic or quasiperiodic motion depending on 
$\Lambda $.\ However in addition to this we find that the orbit shape and
symbol sequence can be drastically changed as well.\ We will be able to see
the overall structure of the change when we begin to look at the bottom
region of the Poincar\'{e} maps.

\subsection{Chaotic Orbits}

The chaotic orbits are those in which the hex-particle wanders between $A$
motion and $B$ motion in a seemingly irregular fashion.\ On the Poincar\'{e}
section these orbits appear as densely filled regions. These orbits
eventually wander into all areas of the $\left( \rho ,\lambda \right) $
plane allowed by the\ energy constraint, a trait not seen in either the
annuli or pretzel orbits. Areas of chaos can be found in this system at the
transition point between annulus and pretzel orbits, though the size of the
chaotic region seems to be highly dependent on the cosmological constant, as
will be shown later.

\FRAME{ftbpFU}{5.0462in}{3.9375in}{0pt}{\Qcb{Examples of chaotic orbits run
for 200 time steps between the pretzel and annulus regions (top right) and
within the pretzel region (bottom left) along with their corresponding
three-particle trajectories truncated after 60 time steps. \ The sizes of
the regions of chaos are highly dependent on the cosmological constant and
the chaotic oribts in the pretzel regions are only found when $\Lambda >0$.}%
}{\Qlb{chaos_examples}}{chaoticexamples.eps}{\special{language "Scientific
Word";type "GRAPHIC";maintain-aspect-ratio TRUE;display "USEDEF";valid_file
"F";width 5.0462in;height 3.9375in;depth 0pt;original-width
6.6945in;original-height 5.2157in;cropleft "0";croptop "1";cropright
"1";cropbottom "0";filename '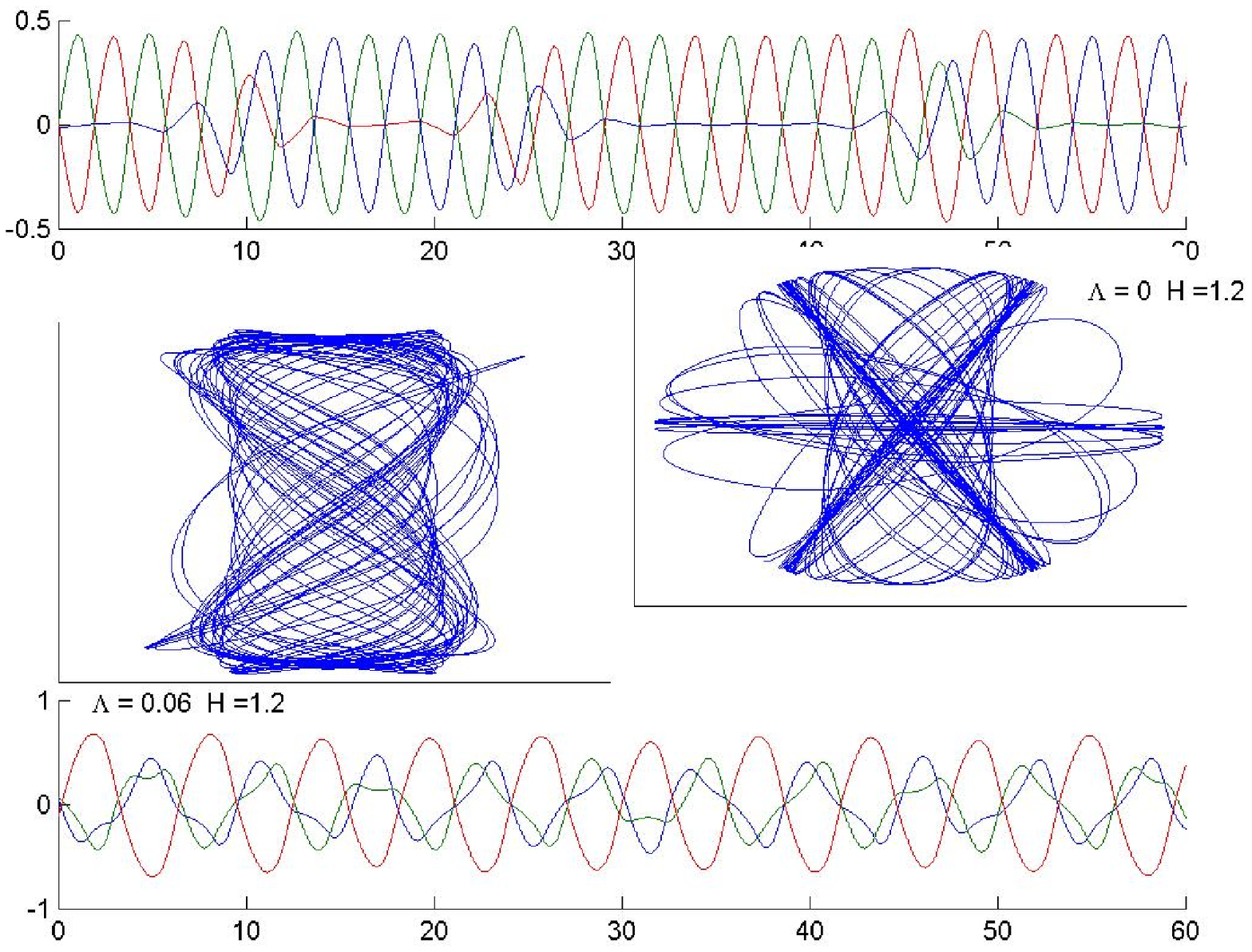';file-properties "XNPEU";}}

Chaotic orbits have been found at all values of $H$ that are numerically
obtainable and have essentially the same basic properties. However these
orbits are generally hard to find at any energy as they are very sensitive
to initial conditions. The top left orbit of figure \ref{chaos_examples}
gives us an example of a chaotic orbit found in between the annulus and
pretzel regions on the Poincar\'{e} section. We can see that the orbit
sometimes shows the characteristic motion of an annulus orbit and sometimes
shows the motion of a pretzel orbit. The three-particle trajectories show us
how the particles switch roles during their quasi-regular motion at
irregular intervals resulting in the chaotic motion.

As with the annulus and pretzel orbits, the chaotic orbits extend farther
out into the $\left( \rho ,\lambda \right) $ plane as $\Lambda $ increases,
and the frequency of the three particle motion increases.\ However, chaotic
orbits are non-periodic by definition, so a change in frequency can never
make this type of orbit periodic.

\FRAME{ftbpFU}{4.6648in}{3.9375in}{0pt}{\Qcb{Chaotic orbits for $H=1.2$
starting with the same FE initial conditions but with different values for $%
\Lambda $ run for 200 time steps (top left $\Lambda =0.06$; top right $%
\Lambda =0$; bottom $\Lambda =-0.175$). \ Not overlapped to emphasize change
in amount of chaotic motion}}{\Qlb{chaos_transfer}}{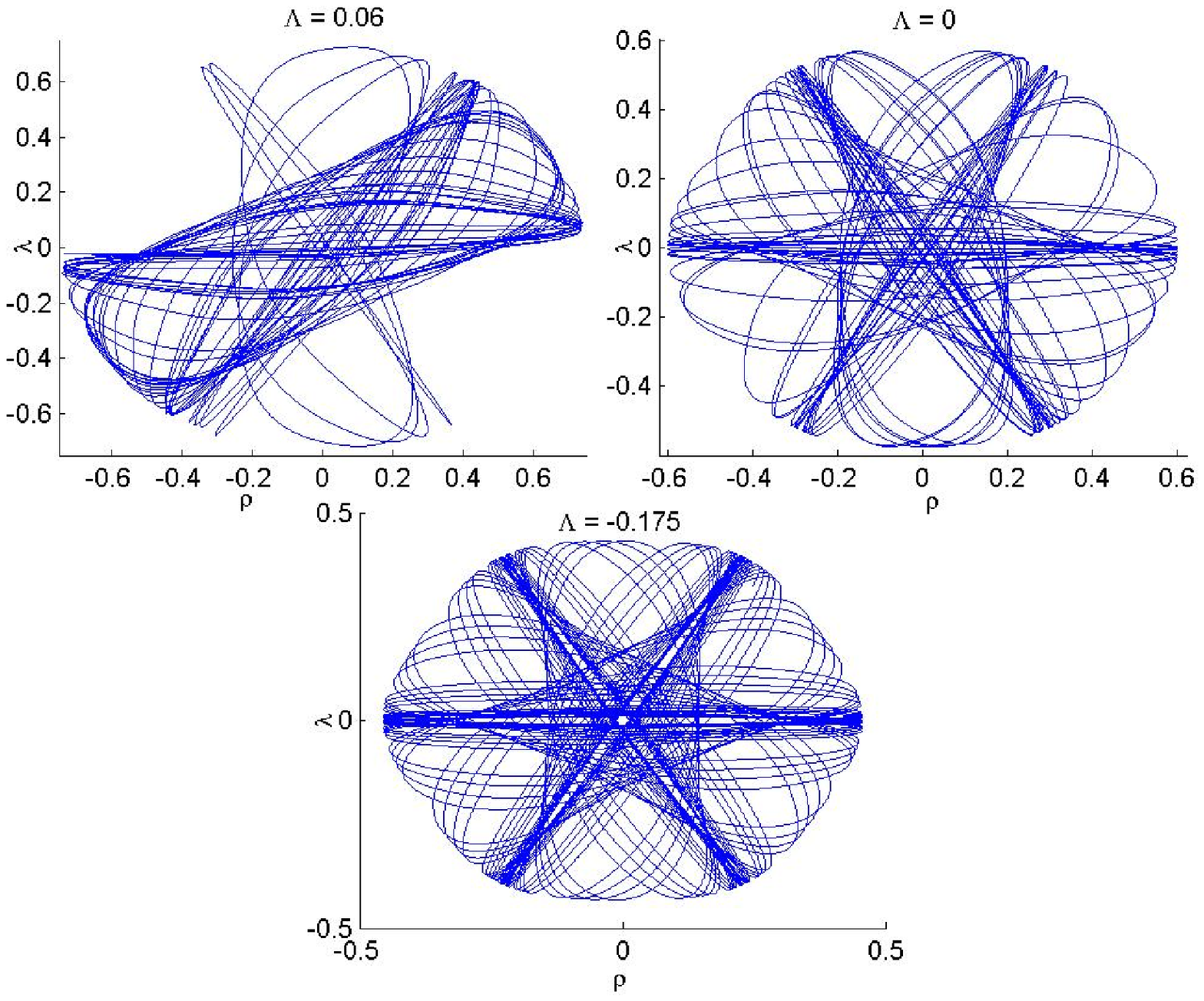}{%
\special{language "Scientific Word";type "GRAPHIC";maintain-aspect-ratio
TRUE;display "USEDEF";valid_file "F";width 4.6648in;height 3.9375in;depth
0pt;original-width 6.5959in;original-height 5.5616in;cropleft "0";croptop
"1";cropright "1";cropbottom "0";filename 'ChaoticTransfer.eps';file-properties "XNPEU";}}

The most significant change to the chaotic orbits at different values of $%
\Lambda $\ is that as $\Lambda $ becomes negative, the initial conditions
that resulted in a chaotic orbit will generally result in either an annulus
or pretzel orbit. Conversely, we have found more initial conditions
resulting in chaotic orbits when $\Lambda $\ is positive.\ Overall there
seems to be a strong positive correlation between the value of the
cosmological constant and the size of the subset of initial conditions that
result in chaotic orbits. This trend is illustrated by figure (\ref%
{chaos_transfer}) where for the same FE initial conditions are run with
different values of $\Lambda $. \ The initial condition that results in
chaotic orbits for $\Lambda \geq 0$\ instead result in a regular annulus
orbit (in this case) for $\Lambda <0$.

In addition, we have also found separate regions of chaos within the pretzel
region of the Poincar\'{e} map when $\Lambda $ is positive. For certain
initial conditions resulting in pretzel orbits when $\Lambda =0$, when we
increase $\Lambda $\ we find that the hex-particle begins to irregularly
transverse increasingly larger regions of the $\left( \rho ,\lambda \right) $%
\ plane. \ However, these chaotic pretzel orbits do not cover the entire $%
\left( \rho ,\lambda \right) $\ plane as do the chaotic orbits in between
the annulus and pretzel regions. \ An example of this type of orbit is shown
on the bottom left of figure (\ref{chaos_examples}). \ In these chaotic
pretzel orbits -- unlike the previously described chaotic orbits where all
three particles switch positions at irregular intervals -- we can see from
the three-particle motion that while two of the particles orbit each other
closely in a bound subsystem, the third particle shows fairly regular
motion. The chaotic motion of the hex-particle results from irregular
motions in the two-particle subsystem.

\section{Poincar\'{e} Plots}

We now consider the Poincar\'{e} sections for this system. These are
constructed by plotting the square of the angular momentum ($p_{\theta }^{2}$%
, labeled as $z$) of the hex-particle against its radial momentum ($p_{r}$,
labeled as $x$) each time it crosses one of the bisectors.\ Since we are
looking only at the case when all three particles have the same mass, all of
the bisectors are equivalent, and so we can plot all of the crossings on the
same surface of section. This allows us to find regions of periodicity,
quasiperiodicity, and chaos.

Since this system is governed by a time-independent Hamiltonian with four
degrees of freedom, the total energy is a constant of the motion and the
phase space is a three-dimensional hypersurface in four dimensions. If an
additional constant of motion exists the system is said to be integrable,
and its trajectories are restricted to two-dimensional surfaces in the
available phase space.\ Since the trajectories can never intersect, that
constraint imposes severe limitations on the types of motion that integrable
systems can exhibit. Trajectories may be periodic\ (repeating themselves
after a finite interval of time) or quasiperiodic. Since the trajectories
are, by definition, comprised by the intersection of two two-dimensional
surfaces, they will always appear as lines or dots for quasiperiodic and
periodic orbits respectively, on the Poincar\'{e} section. This is a sharp
contrast to the case when all orbits can move freely in three dimensions in
a completely nonintegrable system. The extra degree of freedom permits an
orbit to visit all regions of phase space, and the system typically displays
strongly chaotic behaviour.\ Such trajectories appear as filled-in areas on
the Poincar\'{e} map.

When an integrable system is given a sufficiently small perturbation, most
of its orbits remain confined to two-dimensional surfaces. However, it is
possible that small areas of chaos will appear sandwiched between the
remaining two-dimensional surfaces, which can grow as the magnitude of the
perturbation is increased, eventually becoming connected areas on the
Poincare section.\ This phenomenon is called a Kolmogorov-Arnold-Moser (KAM)
transition \cite{KAM}. \ Within these regions, islands of regularity may
remain for quite some time and can have an intricate fractal structure \cite%
{fractal}, but the system will typically become almost fully ergodic for
sufficiently large perturbations \cite{Reichl}.

\FRAME{ftbpFU}{4.849in}{3.9366in}{0pt}{\Qcb{The Poincar\'{e} plot of the
system at $H=1.2$ and $\Lambda =0$. The upper inset shows a close up of the
upper chaotic region where the Poincar\'{e} plot is filled in.\ The lower
inset shows a close up of the structure in the pretzel region.}}{\Qlb{%
poincare_lowh_zerol}}{poincare_lowh_zerol.eps}{\special{language "Scientific
Word";type "GRAPHIC";maintain-aspect-ratio TRUE;display "USEDEF";valid_file
"F";width 4.849in;height 3.9366in;depth 0pt;original-width
6.7914in;original-height 5.5063in;cropleft "0";croptop "1";cropright
"1";cropbottom "0";filename '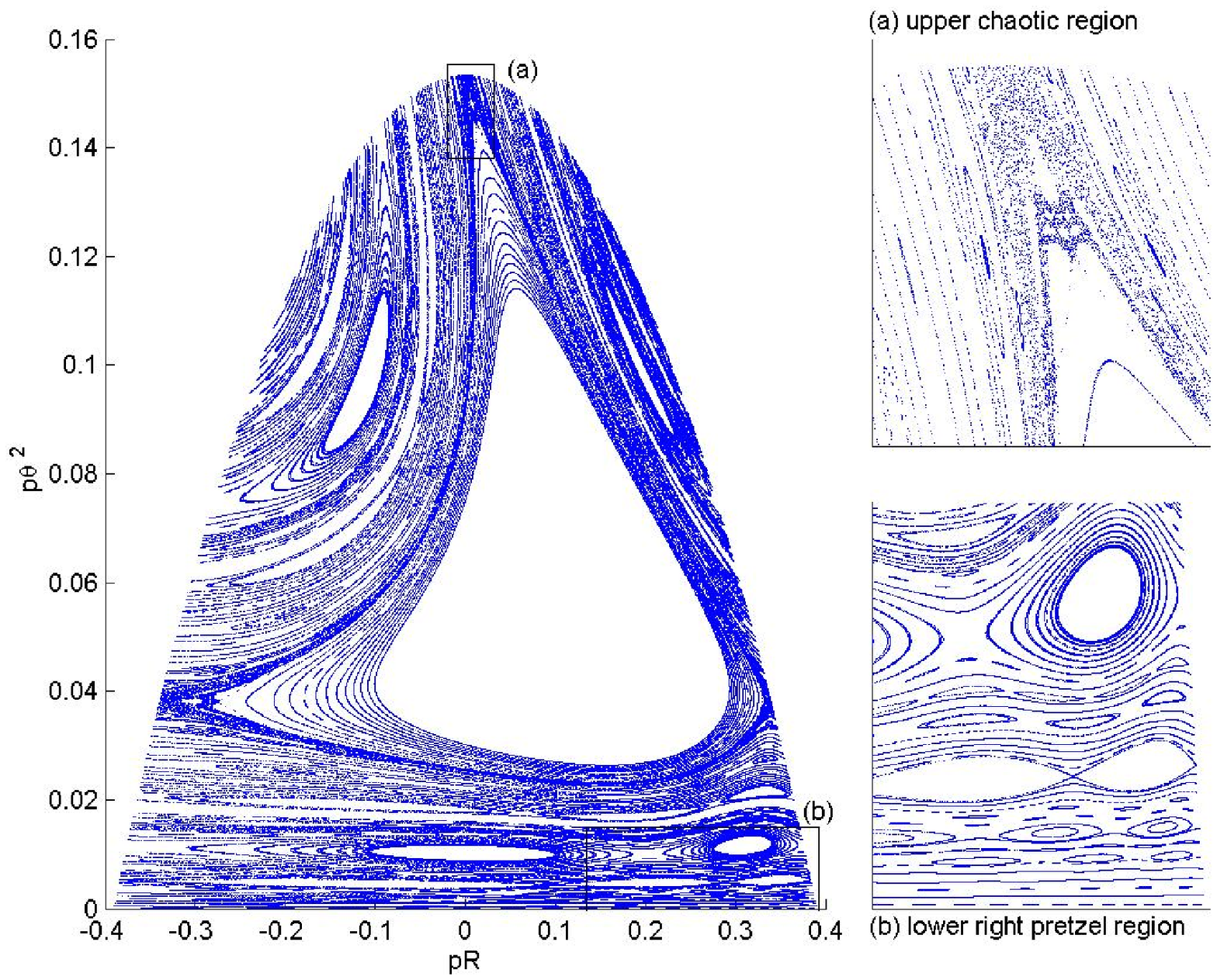';file-properties
"XNPEU";}}

\FRAME{ftbpFU}{4.849in}{3.9366in}{0pt}{\Qcb{The Poincar\'{e} plot for $H=1.2$
and $\Lambda =-0.175$, just above the critical value of -0.18 for that
energy. \ The upper inset gives a close up of the very tiny remaining region
of chaos. \ The lower inset gives a close up of the structure of the lower
right pretzel region.}}{\Qlb{poincare_lowh_negl}}{poincare_lowh_negl.eps}{%
\special{language "Scientific Word";type "GRAPHIC";maintain-aspect-ratio
TRUE;display "USEDEF";valid_file "F";width 4.849in;height 3.9366in;depth
0pt;original-width 6.7914in;original-height 5.5063in;cropleft "0";croptop
"1";cropright "1";cropbottom "0";filename
'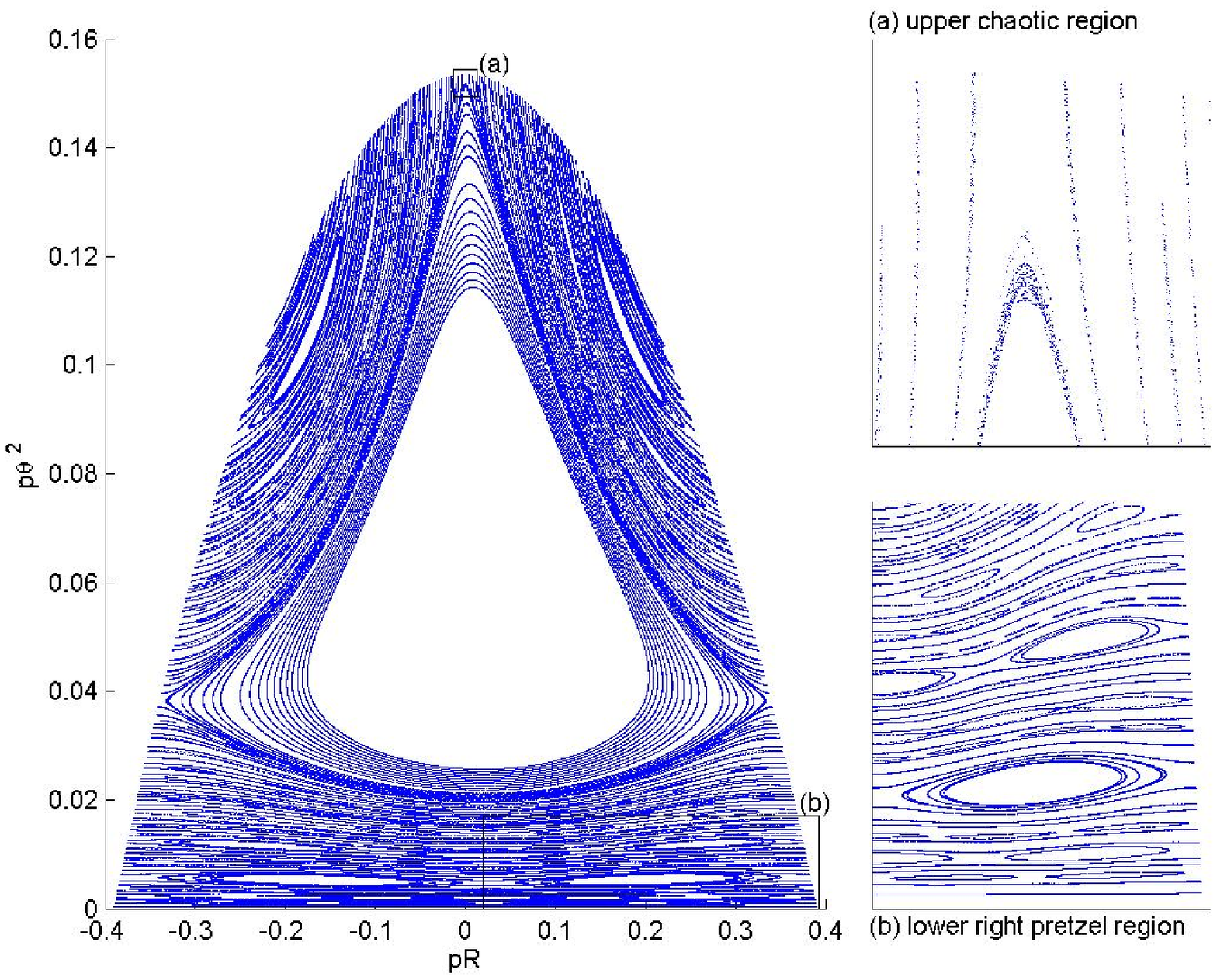';file-properties "XNPEU";}}

\FRAME{ftbpFU}{4.9381in}{3.9366in}{0pt}{\Qcb{The Poincar\'{e} plot for $%
H=1.2 $ and $\Lambda =0.06$, near the highest value we can obtain
numerically. \ The upper and lower insets provide successive closeups of the
lower right pretzel region showing both the regions of pretzel chaos and the
self-similar structure at increasingly small scales.}}{\Qlb{%
poincare_lowh_posl}}{poincare_lowh_posl.eps}{\special{language "Scientific
Word";type "GRAPHIC";maintain-aspect-ratio TRUE;display "USEDEF";valid_file
"F";width 4.9381in;height 3.9366in;depth 0pt;original-width
6.9176in;original-height 5.5063in;cropleft "0";croptop "1";cropright
"1";cropbottom "0";filename '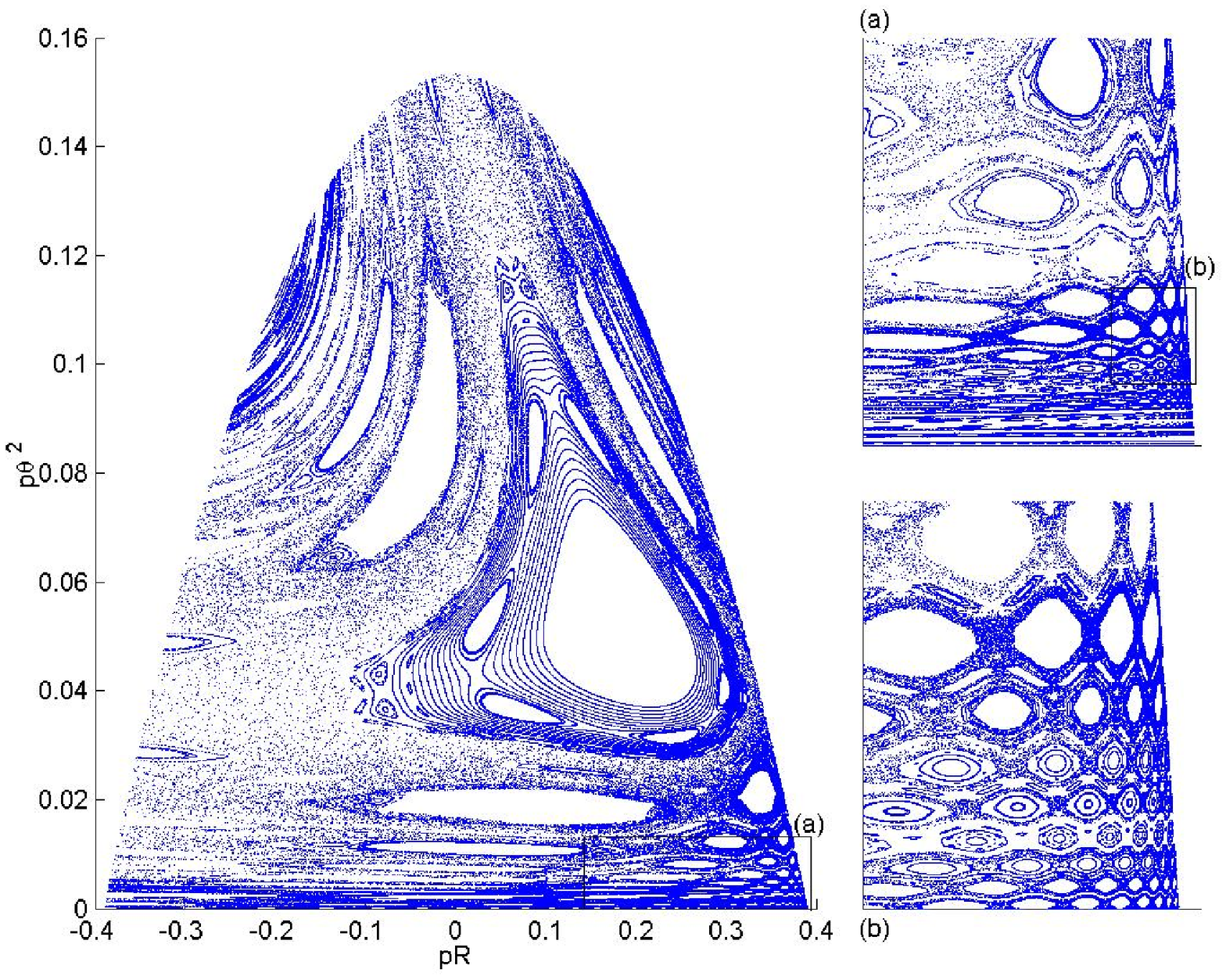';file-properties
"XNPEU";}}

The outer bound of the Poincar\'{e} section is limited by the energy of the
system. The first noticable feature is that it is not symmetric about the $%
p_{r}=0$ axis, but is instead skewed to the right, with the deformation
increasing with $H$. This is at first puzzling; the trajectories of a subset
of the annulus orbits always have positive radial velocities when they
intersect one of the hexagon's edges and the tendency of all annulus orbits
is to have $p_{r}>0$ at the bisectors. However, it occurs because the
Hamiltonian given by (\ref{det-eqn1}) and (\ref{det-eqn3}) is not invariant
under the discrete symmetry $p_{i}\rightarrow -p_{i}$, but is only invariant
under the weaker discrete symmetry $\left( p_{i},\epsilon \right)
\rightarrow \left( -p_{i},-\epsilon \right) $.\ The parameter $\epsilon =\pm
1$ is a discrete constant of integration that is a measure of the flow of
time of the gravitational field relative to the particle momenta.\ In our
study we have chosen $\epsilon =+1$ throughout, which has the effect of
making the principal features of the Poincar\'{e} plot ``squashed'' towards
the lower right-hand side of the figure. If we had chosen $\epsilon =-1$
this deformation would be towards the lower left side. This is reminiscent
of the situation for two particles, in which the gravitational coupling to
the kinetic energy of the particles causes a distorition of the trajectory
from an otherwise symmetric pattern \cite{2bd,2bdchglo} becoming more
pronounced as $H$ increases.

The general structure of the Poincar\'{e} section can be seen in figure (\ref%
{poincare_lowh_zerol}). There is a fixed point just right of the center of
the plot surrounded by triangular rings. These near-integrable curves
correspond to the annulus orbits for that energy. All of the possible
annulus orbits are contained in the largest triangular ring surrounding this
region. Its boundary contains a thin region of chaos and outside of that, at
the bottom and top corners, we have the regions corresponding to the pretzel
orbits.

The structures in the lower part and upper corners of the Poincar\'{e}
section is extremely complicated.\ We find a self similarity in these
regions with a series of circles bounding the quasiperiodic near-integrable
regions repeating themselves on increasingly smaller scales. We find that
each of these circular patterns corresponds to a family of pretzel orbits.
The two largest circles just below the annulus region correspond to the
boomerang-shaped orbits $\left( \overline{AB^{3}}\right) $. The next set of
three circles corresponds to the sequence $\left( \overline{A^{2}B^{3}}%
\right) $, and so on. We see a collection of crescents between these sets of
circles and they correspond to the sequences $\overline{AB^{3}A^{2}B^{3}}$, $%
\overline{AB^{3}AB^{3}A^{2}B^{3}}$, and so on. Inside these circles there is
actually a continuum of possible circles with a diameter that depends upon
the initial conditions. These circles are centered around a single point
that corresponds to the periodic orbit that these orbits approach.

\FRAME{ftbpFU}{5.2486in}{3.9384in}{0pt}{\Qcb{The Poincar\'{e} plot of the
system with $H=1.5$ and $\Lambda =0$. The upper inset provides a closeup of
the upper chaotic region while the lower inset provides a closeup of the
lower right pretzel region.}}{\Qlb{poincare_midh_zerol}}{%
poincare_midh_zerol.eps}{\special{language "Scientific Word";type
"GRAPHIC";maintain-aspect-ratio TRUE;display "USEDEF";valid_file "F";width
5.2486in;height 3.9384in;depth 0pt;original-width 7.9442in;original-height
5.9525in;cropleft "0";croptop "1";cropright "1";cropbottom "0";filename
'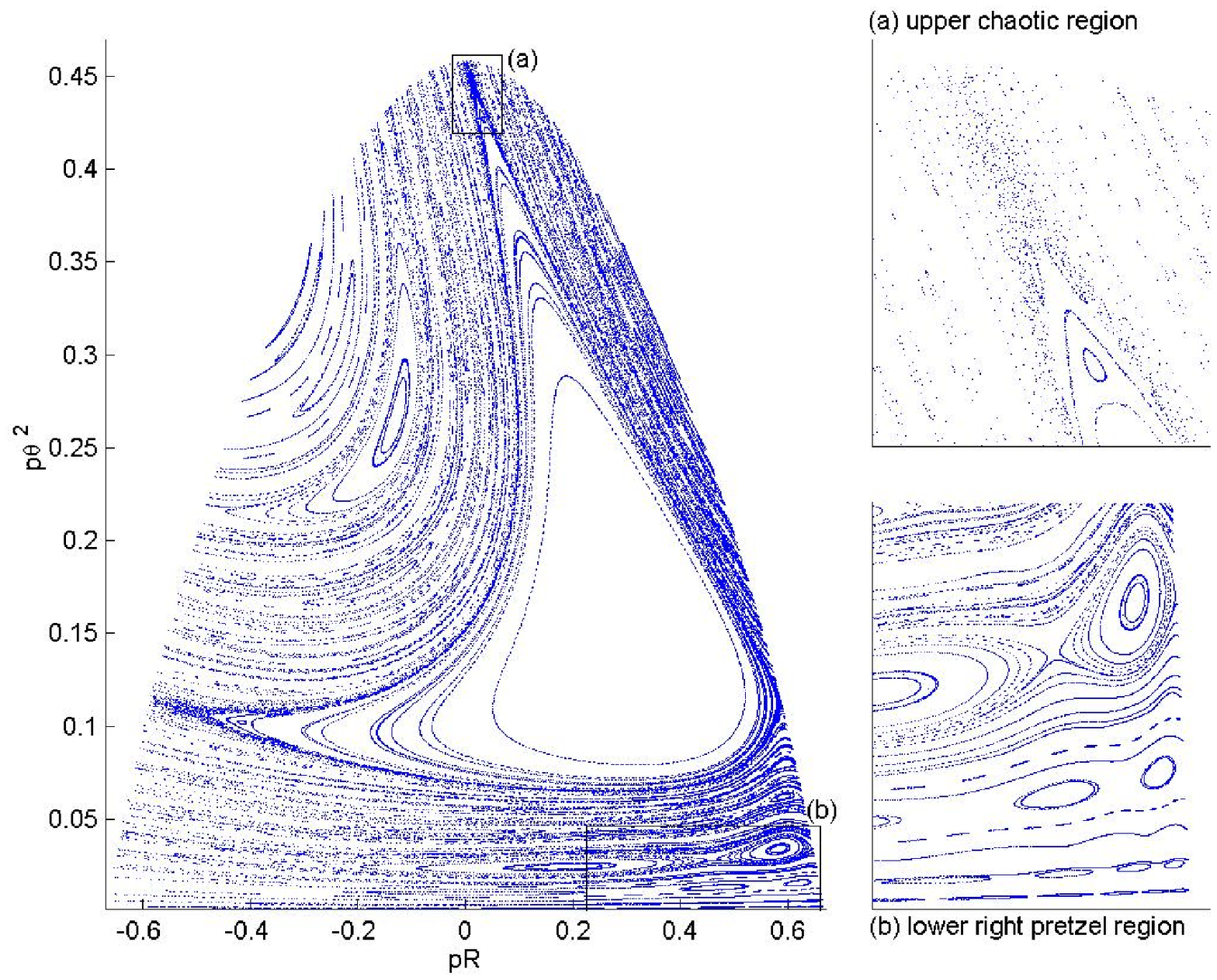';file-properties "XNPEU";}}

\FRAME{ftbpFU}{5.2486in}{3.9384in}{0pt}{\Qcb{The Poincar\'{e} plot for the
system at $H=1.5$ and $\Lambda =-0.25$, just above the critical value at
that energy. \ The upper inset provides a closeup of the upper chaotic
region which has essentially dissappeared. \ The lower inset provides a
closeup of the lower right pretzel region.}}{\Qlb{poincare_midh_negl}}{%
poincare_midh_negl.eps}{\special{language "Scientific Word";type
"GRAPHIC";maintain-aspect-ratio TRUE;display "USEDEF";valid_file "F";width
5.2486in;height 3.9384in;depth 0pt;original-width 7.9442in;original-height
5.9525in;cropleft "0";croptop "1";cropright "1";cropbottom "0";filename
'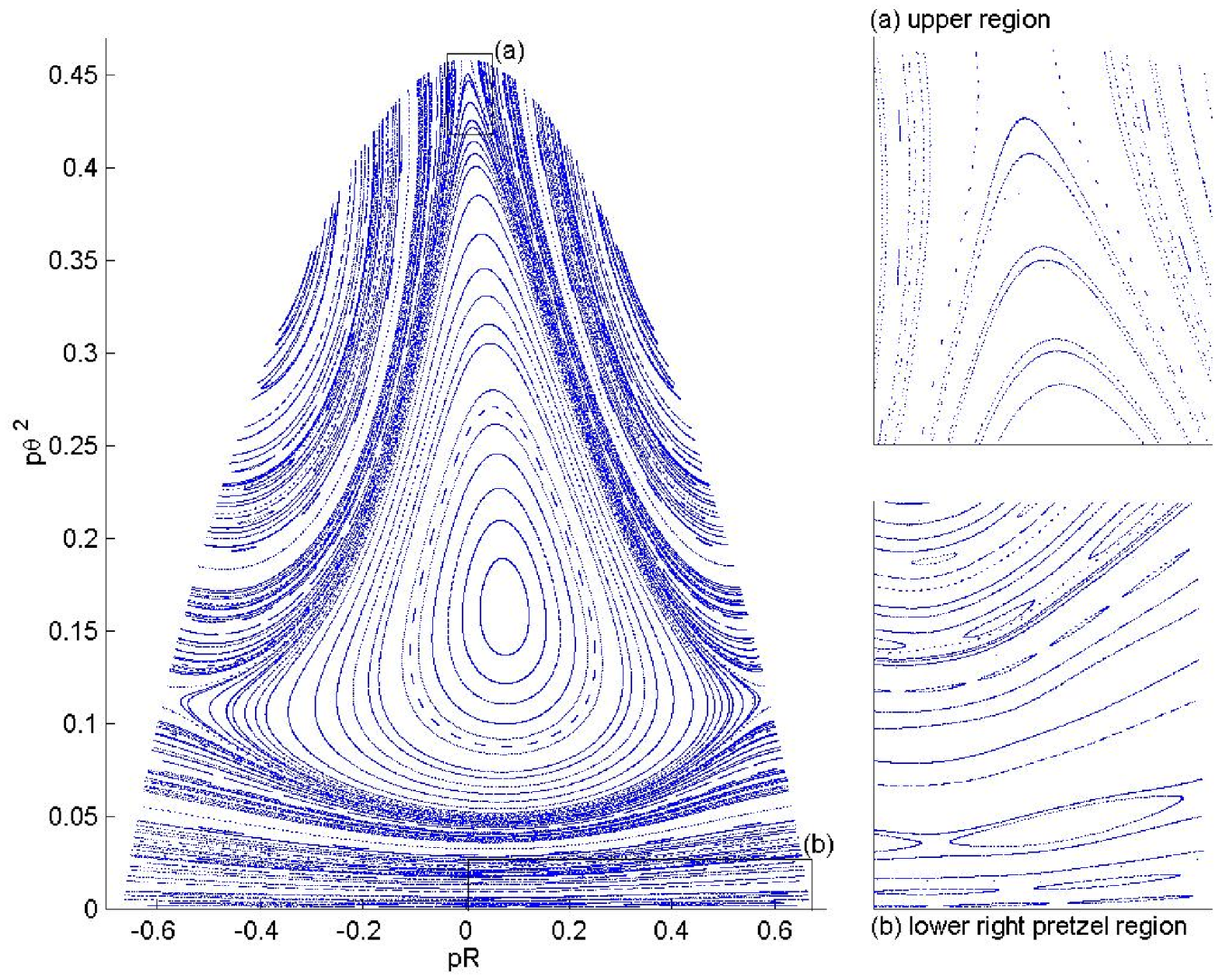';file-properties "XNPEU";}}

\FRAME{ftbpFU}{4.849in}{3.9366in}{0pt}{\Qcb{The Poincar\'{e} plot of the
system at $H=1.5$ and $\Lambda =0.02$, close to the upper limit of $\Lambda $
that is numerically obtainable.}}{\Qlb{poincare_midh_posl}}{%
poincare_midh_posl.eps}{\special{language "Scientific Word";type
"GRAPHIC";maintain-aspect-ratio TRUE;display "USEDEF";valid_file "F";width
4.849in;height 3.9366in;depth 0pt;original-width 6.7914in;original-height
5.5063in;cropleft "0";croptop "1";cropright "1";cropbottom "0";filename
'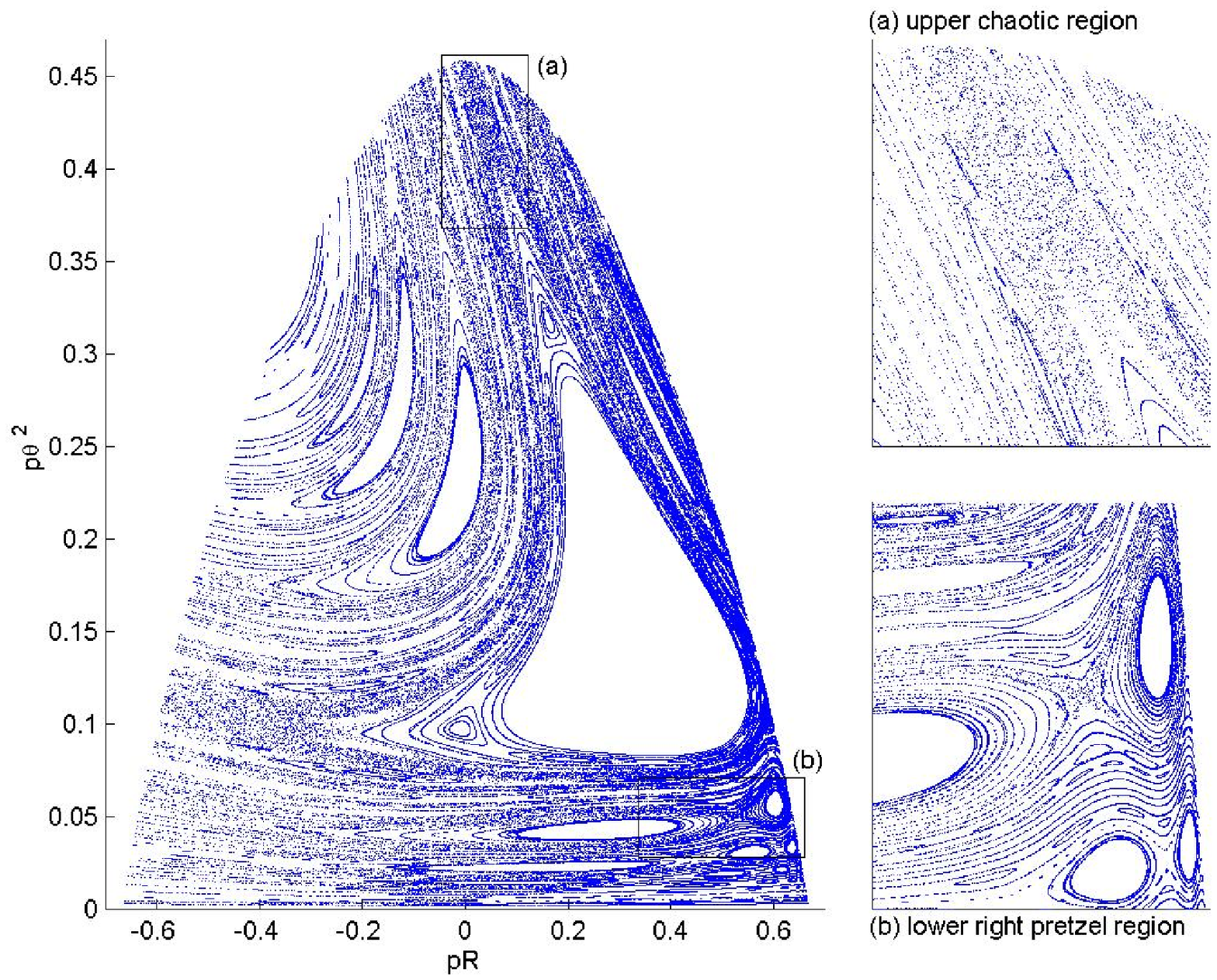';file-properties "XNPEU";}}

As $H$ increases, we remarkably do not find a breakdown from regular to
chaotic motion in this system. As described before we see the plot skewed
farther to the bottom right with larger $H$ but the topology of the system
seems to remain the same. For instance in the lower region of the Poincar%
\'{e} map we see the same patterns of series of circles at all different
energies with no sizeable connected areas of chaos.\ We see the
characteristic regions of chaos appearing in the regions between the annulus
region and the pretzel regions, but we see little change in the relative
sizes of the chaotic regions with $H$. \ 

When we introduce a negative cosmological constant we see some significant
changes in the characteristics of the Poincar\'{e} sections shown in figures
(\ref{poincare_lowh_negl}) and (\ref{poincare_midh_negl}). First, we see
that the asymmetric skewing of the graph to the bottom right is reduced and
``pushed'' back to the center. Specifically as $\Lambda $ approaches its
negative critical value, the Poincar\'{e} section becomes symmetric about
the $p_{r}=0$ line. In the bottom pretzel region we see that the repeating
pattern of circles is continued but again centered on the graph instead of
skewed to the right. This is similar to the low energy Newtonian case
studied in \cite{burnell} where the Poincar\'{e} section is also centered
around $p_{r}=0$. This is commensurate with the interpretation that the
negative critical value of $\Lambda $\ for a given energy corresponds to the
point where all of the energy in the system is vacuum energy, leaving no (ie
very little) energy left for the motion of the particles. \ 

The most suprising feature of the Poincar\'{e} section when $\Lambda <0$ is
the rapid disappearence of all the regions of chaos.\ In the region between
the annulus and pretzel regions we find that once-chaotic orbits split into
purely annulus or purely pretzel class orbits as $\Lambda $ approaches its
negative critical value. We find no evidence of the onset of a KAM breakdown
anywhere in the plot.\ This feature is strikingly different from what we
have seen in the corresponding Newtonian system where chaos is present at
all energies.

In our study of the $\Lambda <0$ case we were able to numerically solve the
equations of motion of the hex-particle for all values of $\Lambda $ up to
the physical limit of the critical negative value where all of the energy of
the system is ``used up'' in the contraction of the spacetime. However, for
the $\Lambda >0$ case, due to the reduction of $V_{rc}$ as $\Lambda $
increases (as previously described) we find that the system that was in a
perturbative regime quickly changes into an intrinsically nonperturbative
system and our numerical simulations break down. This places much greater
restrictions on what values of $\Lambda >0$ we can test, especially at
higher energies. Because of this numerical limit, our study of the $\Lambda
>0$ case is somewhat more limited since we cannot follow the changes to this
system up to a physical limit. However, we have seen the beginnings of a
number of characteristic changes to the Poincar\'{e} section as $\Lambda $
increases, as shown in figures (\ref{poincare_lowh_posl}) and (\ref%
{poincare_midh_posl}).

First, many features are skewed farther to the right as $\Lambda $
increases. Furthermore many of the features in the bottom pretzel region
seem to be skewed up and to the right instead of down. Below the two large
circles in the pretzel region we again see the repeating pattern of circles
very well defined.

We also find that the chaotic region in between the boundary of the annulus
and pretzel regions is also greatly increased, both in the corners of the
boundary and along the sides. These characteristics are found at all
numerically obtainable values of $H$, though for larger $H$, as described
before we cannot increase $\Lambda $ as much before the system becomes
non-integrable.

\FRAME{ftbpFU}{6.5241in}{4.8879in}{0pt}{\Qcb{A closeup of the pretzel region
of the Poincar\'{e} plots with $H=1.2$ for increasing values of $\Lambda $.
The diagrams are all of the same part of the section. \ We see clear
evidence of KAM breakdown as $\Lambda $ increases.}}{\Qlb{kam_break}}{kambreak.eps}{\special{language "Scientific Word";type
"GRAPHIC";maintain-aspect-ratio TRUE;display "USEDEF";valid_file "F";width
6.5241in;height 4.8879in;depth 0pt;original-width 7.9442in;original-height
5.9525in;cropleft "0";croptop "1";cropright "1";cropbottom "0";filename '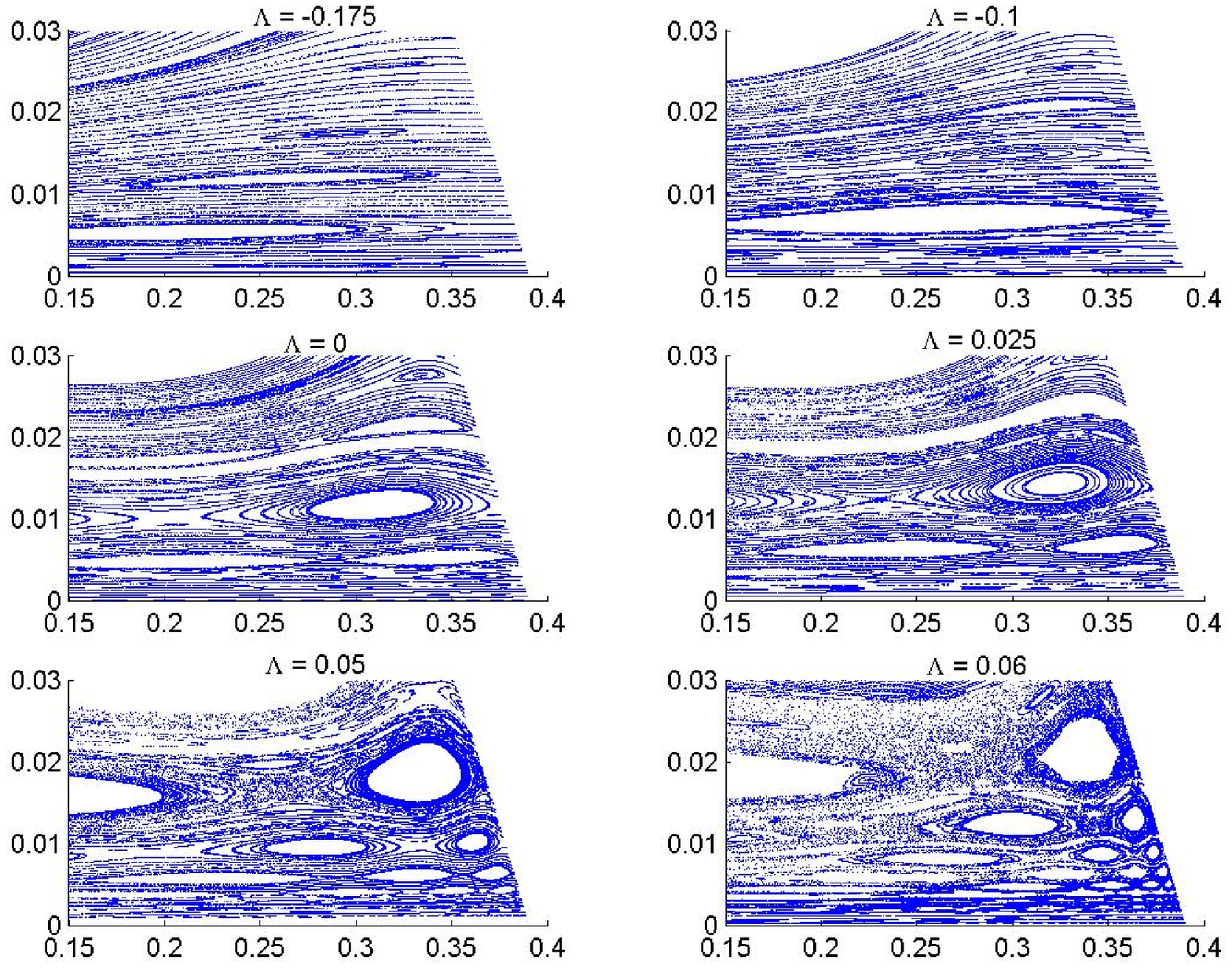';file-properties "XNPEU";}}

Furthermore, as $\Lambda $ increases we at first see the onset of KAM
breakdown in the pretzel region, followed by the appearence of major regions
of chaos. This transition is illustrated in figure (\ref{kam_break}). For $%
H=1.2$, at $\Lambda =0.025$ we already begin to see some of the lines widen
and small regions of chaos appear between the groups of ellipses. In these
narrow regions the hex-particle seems to switch between orbits with an
infinitely repeating symbol sequence (which correspond to the groups of
ellipses) and orbits which undergo additional $A$ type motions in
quasiperiodic intervals (corresponding to the wavy lines) at seemingly
irregular intervals. It is these regions that result in the chaotic pretzel
orbits mentioned before. However, at this point most of the orbits still
show very regular motions. As $\Lambda $ increases to $0.05$ we see these
new regions of chaos expand around the groups of ellipses. Once we increase $%
\Lambda $ to $0.06$ most of the lower region of the Poincar\'{e} section
becomes chaotic, with only a few non-connected regions of regular motion
remaining. This is in sharp contrast with the behaviour of the system as $%
\Lambda $ becomes negative, in which no evidence of any KAM breakdown is
apparent.

\FRAME{ftbpFU}{6.0407in}{5.0488in}{0pt}{\Qcb{Regions of the Poincar\'{e}
section at $H=1.8$ and $\Lambda =0$ for the top graphs and $\Lambda =0.0085$
for the bottom graphs. The graphs on the left show closeups of the upper
chaotic region and the graphs on the right show closeups of the lower
pretzel region.}}{\Qlb{High H KAM}}{highhkam.eps}{\special{language
"Scientific Word";type "GRAPHIC";maintain-aspect-ratio TRUE;display
"USEDEF";valid_file "F";width 6.0407in;height 5.0488in;depth
0pt;original-width 6.6815in;original-height 5.5789in;cropleft "0";croptop
"1";cropright "1";cropbottom "0";filename '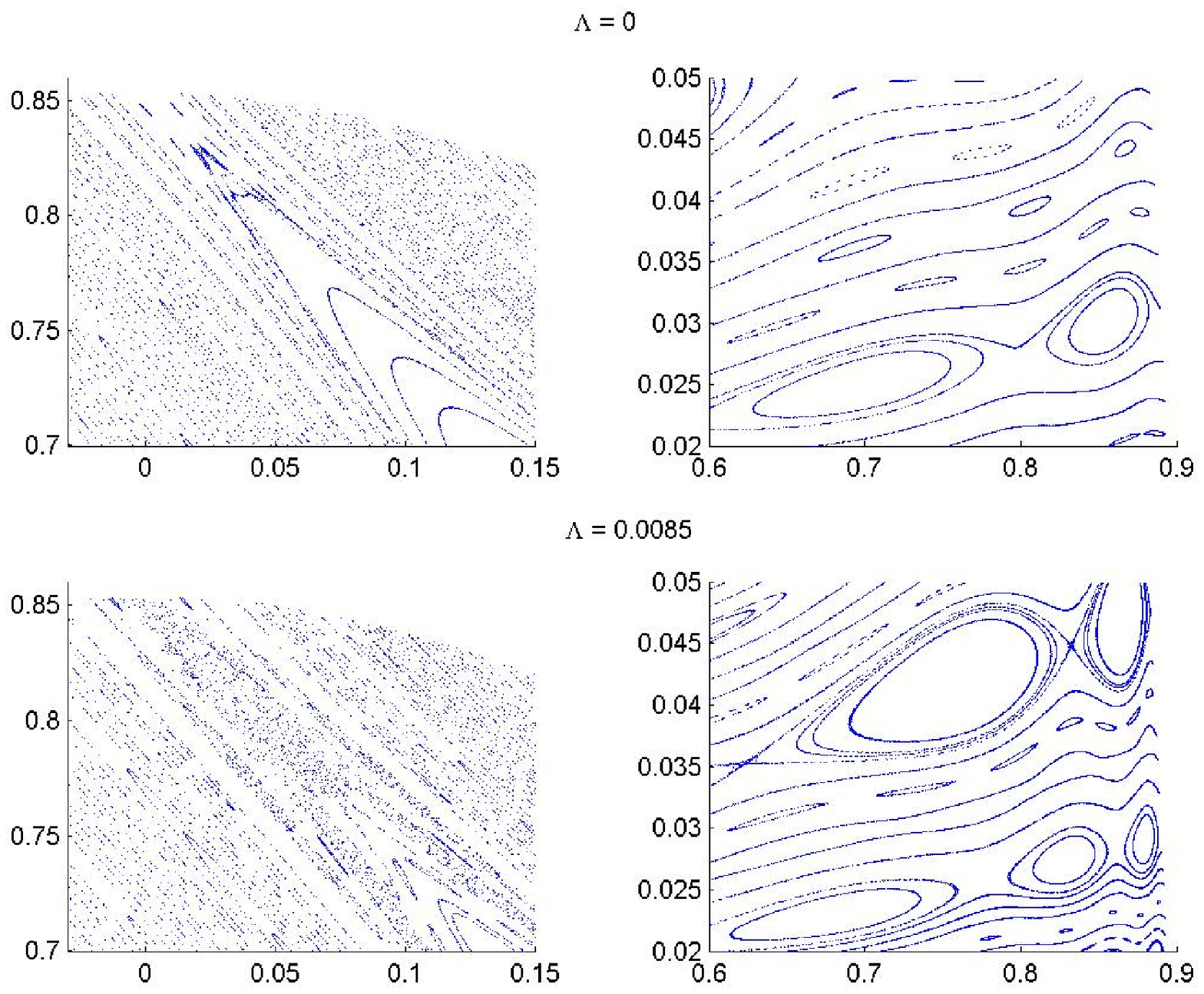';file-properties
"XNPEU";}}

Unfortunately, due to our previously mentioned limitations on studying
systems with large cosmological constants, for systems with higher energies
we have not been able to numerically investigate systems with as large a
value of $\Lambda $, so we have not been able to observe this full KAM
breakdown at all energies. However, for lower values of $\Lambda $ we have
observed the same initial trends of the lines between the series of ellipses
begin to thicken as shown in the lower right insert of figure (\ref%
{poincare_midh_posl}).\ Figure (\ref{High H KAM}) shows another example of
these trends at higher energies. As $\Lambda $\ is increased, we again see
an increase in the area of the upper chaotic regions and the appearance of
very small regions of chaos in between the series of ellipses. This suggests
that the preliminary stages of KAM breakdown occur at these higher energies
as well.

\bigskip

\section{Discussion}

Our investigation of the cosmological 3-body problem has revealed a number
of interesting features. We find in general that the presence of a
cosmological constant significantly modifies the chaotic properties of the
relativistic 3-body system, and this in markedly different ways depending on
its sign.

For a negative cosmological constant we find that there is a rapid decrease
in the amount of chaos for all values of $H$ that we were able to
investigate. The size of the chaotic regions in the Poincar\'{e} plot are
even smaller than in its non-relativistic counterpart, despite the high
degree of non-linearity in the cosmological system. Indeed, as the
cosmological constant approaches its negative critical value (defined in \ref%
{eqn-negcrit}), the chaotic regions nearly vanish. We conjecture that this
occurs for arbitrarily large $H$, motivated primarily by our observation
that the area of the chaotic regions in the Poincar\'{e} section seems to be
roughly proportional to $\left| \frac{\Lambda }{\Lambda _{negcrit}}\right| $
for all energies we were able to numerically investigate.

Conversely, we find an increase in the area of chaotic regions in the Poincar%
\'{e} section when the cosmological constant is positive, both in the
regions between the annulus and pretzel orbits, and within the regions
corresponding to the pretzel orbits. Though we were unable to investigate
very large cosmological constants for systems with higher energies, at
higher energies, even for the reasonably small positive cosmological we were
able to study, we observed the lines in the pretzel regions of the Poincar%
\'{e} section starting to thicken and very small regions of chaos appear
between the groups of ellipses , reminicent of the preliminary stages of KAM
breakdown. We therefore conjecture that this increase in chaos occurs for
positive $\Lambda $ at all energies.

We close with some comments on future work. The unequal mass case remains to
be explored. While we expect that the general feature of chaos
increasing/decreasing with positive/negative $\Lambda $ will still be
present, there could be a number of surprising features in the details
relative to the $\Lambda =0$ case \cite{Justin}. However our primary
limitation has been that of exploring energies that are below the critical
value of the potential. This same limitation was present in previous studies
of the $\Lambda =0$ 3-body system \cite{burnell,Justin} -- to move beyond it
will require employing a time parameter that is not coordinate time, as well
as more sophisticated numerical algorithms that avoid instabilities we
encountered at higher energies.

\bigskip

\section*{Acknowledgements}

This work was supported by the Natural Sciences and Engineering Research
Council of Canada.

\end{document}